# High-Resolution In-situ Synchrotron X-ray Studies of Inorganic Perovskite CsPbBr₃: New Symmetry Assignments and Structural Phase Transitions

*Sizhan Liu  Alexander R. DeFilippo  Mahalingam Balasubramanian  Zhenxian Liu  SuYin Grass Wang  Yu-Sheng Chen  Stella Chariton  Vitali Prakapenka  Xiangpeng Luo  Liuyan Zhao  Jovan San Martin  Yixiong Lin  Yong Yan  Sanjit K. Ghose  Trevor A. Tyson\**

Dr. Sizhan Liu, Alexander R. DeFilippo, Prof. Trevor A. Tyson
Department of Physics, New Jersey Institute of Technology, Newark, NJ 07102, USA
E-mail: tyson@njit.edu
Dr. Mahalingam Balasubramanian
Advanced Photon Source, Argonne National Laboratory, Argonne, IL 60439, USA
Dr. Zhenxian Liu
Department of Physics, University of Illinois at Chicago, IL 60607-7059, USA
Dr. SuYin Grass Wang, Dr. Yu-Sheng Chen, Dr. Stella Chariton, Dr. Vitali Prakapenka
Center for Advanced Radiation Sources, University of Chicago, Argonne, IL 60439, USA
Xiangpeng Luo and Prof. Liuyan Zhao
Department of Physics, University of Michigan, Ann Arbor, MI 48109-1040, USA
Jovan San Martin, Dr. Yixiong Lin, Prof. Yong Yan
Department of Chemistry and Biochemistry, San Diego State University, San Diego, CA 92182, USA
Dr. Sanjit K. Ghose
National Synchrotron Light Source II, Brookhaven National Laboratory, Upton, NY 11973, USA

Perovskite photovoltaic ABX₃ systems are being studied due to their high energy-conversion efficiencies with current emphasis placed on pure inorganic systems. In this work, synchrotron single-crystal diffraction measurements combined with second harmonic generation measurements reveal the absence of inversion symmetry below room temperature in CsPbBr₃. Local structural analysis by pair distribution function and X-ray absorption fine structure methods are performed to ascertain the local ordering, atomic pair correlations, and phase evolution in a broad range of temperatures. The currently accepted space group assignments for CsPbBr₃ are found to be incorrect in a manner that profoundly impacts physical properties. New assignments are obtained for the bulk structure: $Im\bar{3}$ (above ∼ 410 K), $P2_1/m$ (between ∼ 300 K and ∼ 410 K), and the polar group $Pm$ (below ∼ 300 K), respectively. The newly observed structural distortions exist in the bulk structure consistent with the expectation of previous photoluminescence and Raman measurements. High-pressure measurements reveal multiple low-pressure phases, one of which exists as a metastable phase at ambient pressure. This work should help guide research in the perovskite photovoltaic community to better control the structure under operational conditions and further improve transport and optical properties. Accepted by Advanced Science, June 2021

## 1 Introduction

Hybrid perovskite photovoltaic ABX₃ systems have been systematically studied for the past decades due to their high energy-conversion efficiencies (in excess of ∼ 25% [1, 2, 3]. In these systems, the A site is occupied by an organic cation, the B site is occupied by Pb, and the X site is occupied by halides (Cl, Br, or I). The organic component at the A site with charge +1 is believed to be the Achilles' heel accounting for the degradation of the hybrid perovskite system at high temperature and in the presence of moisture. Efforts to mitigate this weakness of the hybrid systems are being aggressively pursued. All-inorganic analogs of the hybrid system with the A site occupied by alkali atoms, such as Cs or Rb, are being investigated [4, 5, 6, 7, 8]. These systems have high stability and high open-circuit voltages. Of particular interest is CsPbBr₃ which is known to have a high carrier mobility and a large carrier diffusion length [9, 10, 11]. Understanding the basic physical principles underlying the exceptional properties under operational conditions of this material requires a detailed determination of the crystal structure.

Early experiments by Moller [12, 13] and Cola *et al.* [14] found a monoclinic cell at room temperature while Hirotsu *et al.* [15] assigned a *Pnma* orthorhombic cell based on the neutron diffraction method.

Specifically, Hirotsu *et al.* observed systematic absence violations and superlattice reflections in this orthorhombic cell but excluded them due to their weak intensities. The superlattice reflection intensities were found to exhibit changes at the bulk transition temperatures indicating that the corresponding planes are an integral part of the $CsPbBr_3$ crystal lattice [15].

In the most recent assessments focusing on photovoltaic properties, the crystal structure of $CsPbBr_3$ was studied mainly by powder diffraction and laboratory diffraction methods [9, 10, 11, 16] utilizing the early assigned space groups by Hirotsu *et al*. These methods are known to have difficulty in detecting very weak reflections which may lead to the assignment of wrong space groups. The early work suggested a second-order change from an orthorhombic to a tetragonal space group at 361 K with increasing temperature, indicating a $Pnma$ (#62, centrosymmetric) to $P4/mbm$ (#127, centrosymmetric) space group change. At 403 K, a first-order transition from $P4/mbm$ to the $Pm\bar{3}m$ (#221, centrosymmetric) space group was found. It is noted that all of these space groups are centrosymmetric. In terms of the simple perovskite cell with edge-length $a_p \sim 5.8$ Å, these transitions correspond to changes in the unit cell volume from $\sqrt{2}a_p \times \sqrt{2}a_p \times 2a_p$ to $\sqrt{2}a_p \times \sqrt{2}a_p \times a_p$, and from $\sqrt{2}a_p \times \sqrt{2}a_p \times a_p$ to $a_p \times a_p \times a_p$, respectively, as the temperature is increased. Based on the early symmetry assignments, the results were refined more recently in powder diffraction measurements [17], nuclear magnetic resonance, and nuclear quadrupole resonance studies [18]. However, no independent structural symmetry assignments were performed on single crystals to determine the space group symmetries unambiguously. At the level needed for accurate theoretical models for electron transport, thermal properties, and to develop accurate atomic potentials, these details have been lacking in the literature.

Recent theoretical and experimental physical property determinations have been found to be inconsistent with the established space groups. Density functional theory (DFT) based lattice dynamics calculations exploring the phonon stable models of $ABX_3$ inorganic alkali halide systems at low temperature reveal instabilities in the current cubic, tetragonal, and orthorhombic phase of $CsPbBr_3$ [19]. The stable phase is predicted to be monoclinic $P2_1/m$. In the specific class of Pb-based alkali halides $APbX_3$, DFT modeling reveals distortions as classic double-well potentials relative to the cubic phase [20]. The results indicate collective ferroelectric polarization which may be observable on the nanoscale. In quite recent experiments, quantum dots of $CsPbBr_3$ (cubic-shaped nano-crystals with $\sim 5$ nm edge lengths) were found to exhibit a finite electric polarization. The value of the saturation polarization at 298 K and 77 K were found to be $\sim 0.018$ $\mu C/cm^2$ and $\sim 0.25$ $\mu C/cm^2$, respectively [21]. A significant piezoelectric response has been found in films of this material ($\sim 40$ nm to $\sim 260$ nm thick) [22]. Room temperature remnant polarization values up to $\sim 0.03$ $\mu C/cm^2$ were observed. These recent studies suggest the existence of non-centrosymmetric structures and new symmetry assignments in bulk $CsPbBr_3$ at room temperature and below. In the work presented here, independent structural measurements were conducted on micron-scale as-grown single crystals to assess the structure of this material on multiple length scales.

## 2    Results and Discussion

To determine the appropriate symmetries and phase transitions in this material, high-quality single crystals (orange phase) were synthesized and studied. An as-grown single crystal with $\sim 50$ $\mu m$ cube edges was utilized for single-crystal diffraction measurements. Second harmonic generation (SHG) measurements were conducted to assess the nature of the crystal symmetry below room temperature. Differential scanning calorimetry (DSC) measurements between 300 and 700 K indicate that both transitions in this region are first-order with the transition near 350 K having a smaller $\Delta H$, relative to that at $\sim 410$ K. Single-crystal X-ray diffraction (XRD) measurements were conducted between 100 and 450 K, utilizing a detector with a high dynamic range to detect both weak and strong reflections.

Our high-resolution synchrotron-based single-crystal diffraction measurements reveal that between $\sim 450$ K and 410 K, the space group is cubic $Im\bar{3}$ (#204, centrosymmetric). It undergoes a first-order cubic-to-monoclinic phase transition at $\sim 410$ K. Between $\sim 410$ and 300 K, the $P2_1/m$ (#11, centrosymmetric) monoclinic space group is maintained. A first-order isostructural transition within the $P2_1/m$ space group is observed at $\sim 350$ K. As temperature goes below $\sim 300$ K, a second-order phase transformation is identified. Below 300 K, the weak superlattice peaks which appear below $\sim 410$ become enhanced in the monoclinic $Pm$ (#6, polar) space group. SHG measurements confirmed the absence of inversion symmetry below 300 K. The currently accepted space groups are found to be incorrect in a manner that profoundly impacts physical properties. The unit cell dimensions, $2a_p \times 2a_p \times 2a_p$, are preserved in the studied temperature range (450 K and below). The previously reported space groups in the literature can be recovered if exclusively the dominant reflections are utilized in structural analysis. Raman scattering measurements between 100 and 830 K and pair distribution function measurements between 10 K and 500 K indicate the presence of an isostructural order-disorder transition near 170 K. High-pressure measurements between 1 atm and 13 GPa indicate the resilience of the material under pressure and reveal a first-order phase transition at $\sim 1$ GPa, and successive continuous transitions near 2 GPa, 6 GPa, and 13 GPa. Low-pressure ($\sim 1$ GPa) room temperature measurements recover the low-temperature properties. Theoretical models of these materials will be more heavily constrained by the utilization of the high-resolution structural data over the broad range of temperatures and pressures provided in this work. The transitions at $\sim 350$ and $\sim 410$ K are found to be first-order. The first-order nature of these successive transitions will impact transport properties if materials are cooled rapidly once heated significantly above room temperature. The details of the experimental and modeling methods are given in the supplementary document. Crystallographic data on the specific phases at representative temperatures are also provided.

Differential scanning calorimetric measurements on $CsPbBr_3$ crystals were used to identify the nature of the observed structural phase transitions. Figure S1 shows the heating and warming curves with transitions evident near 362 K and 402 K. These measurements reveal hysteretic behavior covering the region between 360 K and 365 K and a large peak near the transition at 402 K (with cooling/warming offset), indicating that both transitions are first-order with the lower transition revealing a significantly lower $\Delta H$ in the lower-temperature transition. The weakness of the peak near 360 K in earlier measurements [16] led to its assignment as a second-order transition. As seen below, our X-ray diffraction measurements on single-crystal $CsPbBr_3$ confirmed the nature of both phase transitions as first-order. Discontinuities are seen in the temperature dependence of the lattice parameters and/or Bragg peak intensities at these transitions.

To understand the changes in structure over a broad temperature range, temperature-dependent Raman spectroscopy measurements on single-crystal $CsPbBr_3$ were carried out. The low-energy spectra for limited temperatures have been previously reported [23]. In this work, we present a more complete temperature range to enable the identification of phase transitions. The full Raman spectra can be found in Figure S2. Spectra are systematically shifted in intensity for clarity. Figure 1(a) shows a contour plot of the temperature dependence of the unpolarized Raman spectra from 100 K to 500 K, indicating four transitions as horizontal lines at $\sim 170$ K, 300 K, 350 K, and 410 K. In particular, the 73 $cm^{-1}$ and 79 $cm^{-1}$ modes soften as temperature increases and merge to become a single peak at $\sim 300$ K revealing a second-order phase transition. Note the abrupt changes in the position of these peaks near 350 K. Furthermore, these peaks undergo significant broadening above 410 K. These two modes deserve a detailed analysis due to their significant variation with temperature. Hence, the spectra are fitted to a sum of Lorentzian functions. As shown in Figure 1(b), the black open circles represent the experimental data, the solid red line represents the total fit, and the dashed lines show the individual Lorentzian components. A double-peak structure is shaded in red and blue for the modes at 73 $cm^{-1}$ and 79 $cm^{-1}$, respectively.

Figure 1(c) shows the temperature dependence of the ratio of the fitted peak areas and peak widths for this double-peak structure. With increasing temperature, the Raman scattering intensity ratio of the high-energy to the low-energy peak grows substantially between 100 K and 170 K, while showing a kink at $\sim 170$ K. This is followed by a strong decrease of 79 $cm^{-1}$ mode intensity as temperature increases. The peak area ratio vanishes near 300 K. We note that the cooling and warming curves coincide, indicating no hysteresis. Hence, the phase transition at 170 K is second-order. These transitions near 170 K and 300 K were not previously identified in structural measurements.

DFT simulations indicate that these specific modes involve complex motion of the Cs and Br atoms in which layers of Cs atom exhibit shear-type motion triggered by asymmetric distortion of the $PbBr_6$ polyhedra. Table S1 gives the DFT derived phonon frequencies for the orthorhombic cell while the atomic displacements of representative modes at 32, 56, 73, 79, 137, and 158 $cm^{-1}$ are displayed in Figure S3(b).

We note that, in the case of $CsPbCl_3$, the Raman measurements reveal an order-disorder transition near $\sim 170$ K [24]. Very early nuclear quadrupole magnetic resonance measurements on $CsPbBr_3$ reveal the appearance of an additional line near 167 K, which broadens and disappears for higher temperatures indicating that it is second-order in nature [25]. Pair distribution function (PDF) measurements will be used to explore the nature of the transition near 170 K and the higher temperature transitions. Assessing the space groups and structural symmetry requires detailed high-resolution single-crystal X-ray diffraction measurements.

Accurate structural parameters of bulk $CsPbBr_3$ were derived from detailed synchrotron-based single-crystal diffraction measurements between 100 and 450 K on warming using an as-grown cube-shaped single crystal of edge $\sim 50$ $\mu m$. To observe both strong and weak reflections simultaneously, a detector with a large dynamic range was utilized. Full single-crystal data sets (no symmetry assumptions) were collected in 10 K steps over the temperature range. Systematic exploration of the space groups which best fit the data and account for all observed reflections was conducted.

As indicated above, earlier experimental studies [16, 17, 15] claimed that the high-temperature phase of $CsPbBr_3$ is cubic $Pm\text{-}3m$ structure with a lattice parameter $a_p \sim 5.87$ Å. However, by carefully examining the reciprocal lattice images (for temperatures up to 450 K) obtained from high-resolution single-crystal X-ray diffraction data, half-integers $(h\ k\ l)$ peaks are observed which reveals that the lattice parameters and space groups should be revised. By not accounting for these critical weak reflections, the previously reported models can not properly characterize the properties in this system. We note that although no symmetry assignments were made utilizing all reflections as done here, the presence of these half-integer reflections was reported in early work [12, 13, 14, 15].

In this work, the single-crystal diffraction derived structures are present in Figure 2(a). The space group above 410 K is found to be cubic $Im\overline{3}$ with unit cell volume, $2a_p \times 2a_p \times 2a_p$. Only very weak rotation and tilting can be observed in the cubic structure. Below 410 K, the cubic-to-monoclinic phase transition is characterized by the motions of Br and Cs atoms off high-symmetry positions in the unit cell yielding space group $P2_1/m$ with cell volume, $2a_p \times 2a_p \times 2a_p$. Below 300 K, the structure is in the monoclinic $Pm$ space group. The changes in temperature-dependent lattice parameters also reveal the phase transitions shown in Figures 2(d) and (e).

It should be emphasized that reciprocal space images were examined in detail for all regions of temperature to assign the appropriate space groups. Figure 3(a) shows the single-crystal X-ray diffraction reciprocal lattice image of the $(h\ k\ 0)$ plane at 450 K. The $(h\ k\ l)$ grid corresponds to the previously reported $Pm\overline{3}m$ unit cell with lattice constant $a_p \sim 5.87$ Å. The insets show the 3D intensity of some selected reflections with their asymmetric diffuse scattering background evident. The observation of half-integer $h$ or $k$ values indicates the corresponding lattice constants should be doubled. To systematically compare data for the whole temperature range, the reciprocal lattice precession images for each temperature were calculated based on the same unit cell dimension, $a_p \times a_p \times a_p$. The temperature-dependent intensity map of $(h\ \text{-}2\ 0)$ and $(h\ \text{-}2.5\ 0)$ reflections are given in Figure 3(b) and (c), respectively, which are the selected regions shown in panel (a). The temperature-dependent structural distortion and symmetry changes can be characterized by the intensity variation of the reflections. As an example of the dominant reflections, the intensity of the $(2\ \text{-}2\ 0)$ reflection shows an abrupt change from $\sim 350$ K to 410 K. For the superlattice reflections, the intensity of $(2\ \text{-}2.5\ 0)$ reflection vanishes above $\sim 410$ K, while the $(1.5\ \text{-}2.5\ 0)$ reflection intensity showing a kink at the same temperature. In addition to the change in slope at $\sim 410$ K for the $(2\ \text{-}2\ 0)$ peak intensity in Figure 3(b) , a similar abrupt slope change indicates a transition also occurring at $\sim 350$ K. No change in the space group is found by our structural studies leading to our assignment of an isostructural transition. This type of transition is rare, strictly first-order in nature, and requires a high-order expansion of the Landau free energy [26, 27]. Below $\sim 410$ K, the appearance of the additional reflections reveals the cubic-to-monoclinic phase transition.

Focusing on detailed results of structural refinements, Figure 2(b) shows the quality-of-fit parameters, $R_1$, of the $Pm$, $P2_1/m$, and $Im\overline{3}$ space groups as a function of temperature. In monoclinic cells with beta angles near 90° there is the possibility of pseudomerohedral twining due to rotation of the unit cell by 180° about the $a$ or $c$ axes. Models including this type of twinning were examined. For $Pm$ and $P2_1/m$ space groups, the close squares indicate the R1 fitting parameters which incorporate a pseudomerohedry [28] twin law representing a twofold rotation about the monoclinic $a$ axis, while the open squares are the R1 parameters of racemic twinning for $Pm$ structure and no twinning for $P2_1/m$ structure. The racemic twinning in the $Pm$ structure indicates the existence of a temperature-independent 50/50 distribution of inverted and non-inverted polar domains by volume in the studied temperature range, as expected for an unpolarized sample [29, 30]. The temperature-dependent pseudomerohedry twin domain fraction is given in Figure 2(c). Below 410 K, possible space groups are the non-centrosymmetric $Pm$ or the centrosymmetric $P2_1/m$ space groups. Intensity statistics strongly favored the non-centrosymmetric space group. However, incorporation into the refinement of the pseudomerohedry twin law and a racemic twin fraction significantly improves the R1 parameters between 260 K and 410 K for both models leading to similar R1 values. Even with the lower number of parameters utilized in the $P2_1/m$ model (111 parameters for $P2_1/m$ model and 209 parameters for $Pm$ model), refinement of the $P2_1/m$ model converged to better residual values, 4.4/-4.8 e/Å$^3$, than that of $Pm$ structure, 4.8/-5.3 e/Å$^3$, at 350 K. In addition, a careful survey of the data sets revealed systematic extinctions of ($0k0$) reflections for $k = 2n + 1$ above 300 K. The centrosymmetric monoclinic space group $P2_1/m$ is thus uniquely defined as the structure between 300 K and 410 K. The pseudomerohedry twin domains are formed as a result of the $Pm$ to $P2_1/m$ phase transition. This continuous transition onsets at $\sim$ 260 K and is completed at $\sim$ 300 K. The inset in Figure 2(c) reveals an abrupt change of twin domain fraction at $\sim$ 350 K indicating a first-order phase transition. A comparison of $P2_1/m$ structure and the early reported $P4/mbm$ structure solved from the same data is given in Table S2-B. The maximum displacement is 0.18 Å for Cs and 0.32 Å for Br. Figure S6 shows a good match of the experimental reciprocal space images with the simulated pattern of the $P2_1/m$ structure solution.

In comparison to powder diffraction measurements, the single-crystal diffraction method is more adequate to distinguish between the $P4/mbm$ and $P2_1/m$ space groups or between the $Pm\bar{3}m$ and $Im\bar{3}$ space groups as shown in Figure S7 and S8. In powder diffraction data (see Figure S12-A), weak superlattice peaks overlap with peaks derived from the smaller approximate unit cells especially at a high $2\theta$ angle. As shown in Figure S9, only strong reflections can be captured and utilized with previously proposed unit cell dimensions. While in this work, we utilized all statistically significant reflections for structural refinements. The unfitted reflections in the early work have half-integer ($h\,k\,l$) values based on the old unit cell dimensions. These half-integer reflections are about $10^2$ to $10^4$ times weaker in intensity than the integer ones and can be indexed only with a larger unit cell. Capturing the half-integer reflections requires synchrotron-based XRD measurements and high-quality single-crystal samples. Utilizing the weak reflections exclusively in structural refinement yields the same atomic positions that are refined with the dominant reflections at high temperature as shown in Table S3. The $F_{obs}$ vs. $F_{calc}$ plots for the superlattice peaks and main peaks in $Im$-3 structure are given in Figure S8(f). Thus, the half-integer peaks indeed come from the same structure as the main peaks. Figure S8(g) and S7(h) show a good match of the experimental reciprocal space images with the simulated pattern of the $Im\bar{3}$ structure solution. Accounting for these weak reflections will lower the symmetry by including additional distortions to the previously assumed higher-symmetry structures. The atomic displacement of the newly proposed structures in this work compared to the structures solved from the same data but in early reported space groups are given in Table S2. At 450 K, the $Im\bar{3}$ space group incorporates distortions of 0.11 Å in the Br position away from the high-symmetry special positions in the $Pm\bar{3}m$ structure.

For temperatures below 300 K, as shown in Figure S10, the reciprocal lattice images reveal half-integer reflections as seen in early work [12, 13, 14, 15]. These weak reflections are also $10^2$ to $10^4$ times weaker than the main peaks at high temperatures. Utilization of a subset of the data including only high-intensity reflections will yield the $Pnma$ space group at room temperature as found in previously published works. Significant systematic absence violations are observed in all of the possible orthorhombic space group solutions with this unit cell dimension, $\sqrt{2}a_p \times \sqrt{2}a_p \times 2a_p$. The violations are evident as peaks with intensities that are about $10^2$ lower than the dominant peak intensities. In particular, the currently accepted space group, $Pnma$, has 378 systematic absence violations at 300 K. The list of systematic absence violation are given in Table S4 and Figure S11. On the other hand, careful considerations showed that all these previously unfitted half-integer reflections in the orthorhombic model can indeed be indexed on a primitive monoclinic supercell with cell dimension $\sim 2a_p \times 2a_p \times 2a_p$. These weak reflections as well as the systematic absence violations are the signature of the weak distortions breaking the center of symmetry in $Pnma$ structure model. As shown in Figure S12, the simulated powder patterns of the low-temperature models indicate that these weak reflections are rather critical to distinguish the appropriate space group of the low-temperature phase. This points to the need for single-crystal measurements for complete structural solutions. The structure which accounts for all reflections in the diffraction data for temperatures below 300 K is $Pm$ space group. Figure S13-B shows a good match of the experimental reciprocal space images with the simulated pattern of the $Pm$ crystal structure solution. Compared to the $Pnma$ structure, the main distortion in $Pm$ structure occurs in Cs and Br atoms as shown in Table S2-C. Representative structural parameters from the single crystal structure solutions are given in Tables S5 to S10. Supporting the low-temperature assignments, rotational anisotropy second harmonic generation experiments indicate that the true space group has no inversion center for the low-temperatures phase between 290 K and 190 K.

Rotational anisotropy second harmonic generation (RA-SHG) measurements were carried out to detect the crystal structure characteristics of low-temperature phase below 300 K. It is noted that in our Raman, single-crystal diffraction, and pair distribution function measurements, a centrosymmetric-to-noncentrosymmetric transition occurs at $\sim 300$ K. Consequently, our RA-SHG measurements in this work were conducted at and below 290 K in a cryostat to ensure that the sample was in the low-temperature phase. This contrasts with previous SHG measurements conducted at room temperature with no temperature control [31]. In another recent SHG experiment, the SHG photon energy was well below the band gap of $CsPbBr_3$, resulting in very low SHG conversion efficiency [32]. For this work, the geometry of the RA-SHG experiments is shown in Figure 4(a) where oblique incidence with a small incident angle $\theta$ was used. Incident and reflected light can be selected to be either parallel ($P$) or perpendicular ($S$) to the scattering plane. The 800 nm laser beam was focused on the $ab$ plane of the $CsPbBr_3$ sample, with the full width at half maximum (FWHM) of the focused beam being $\sim 20$ $\mu m$ and the penetration depth of 0.42 $\mu m$ at 800 nm (0.12 $\mu m$ at 400 nm) [33]. The spectrum of the incident fundamental light shown in Figure 4(b) exhibits a central wavelength at 800 nm with a FWHM of 41 nm, which corresponds to the SHG spectrum centered at 400 nm with a FWHM of 15 nm. This fact together with a 400 nm center, 40 nm FWHM bandpass filter before the detector ensures that any detected signal belongs to the SHG instead of the two-photon fluorescence around 520 nm [33] from the sample. The instrument utilized in the experiments is sensitive to SHG signals at levels as small as 10 fW.

The measured RA-SHG signals are given in Figure 4(c). It is observed that the signals in $S_{in}$-$S_{out}$ and $S_{in}$-$P_{out}$ channels taken at 290 K and 190 K have almost identical patterns with slightly different intensity levels. Such features are seen across the whole temperature range from 290 K to 190 K, indicating that there is only one structural phase in this temperature region. Moreover, by calculating the field strengths of the incident fundamental and reflected SHG signal, we found that the average SHG susceptibility tensor magnitude is about 0.1 pm/V which is in the range of typical electric dipole SHG susceptibility values [34] and is about three orders of magnitude larger than those of electric quadrupole [35]. Note that the presence of multiple domains in the measurement volume will spread the radiation emission of the SHG signal over a broader angular range compared to that of the corresponding single domain crystal [36]. Hence, the existence of strong SHG spectra [Figure 4(c), $S_{in}$-$S_{out}$ and $S_{in}$-$P_{out}$ channels] from the electric dipole process demonstrate that the crystal possesses a non-centrosymmetric structure over the temperature range examined from 290 K to 190 K.

To understand the local structure, PDF and X-ray absorption measurements were also conducted. X-ray PDF measurements were conducted between 10 K and 500 K. Fits to the PDF data were conducted over the real space range $2 \leq r \leq 30$ Å compared to the single unit cell averaged parameters explored in the single-crystal diffraction study above. The PDF goodness-of-fit parameter Rw was obtained as a function of temperature for a range of space groups explored in the single-crystal methods. (The $Rw = \left\{ \frac{\Sigma_{i=1}^{N} [G_{Obs}(r_i) - G_{Calc}(r_i)]^2}{\Sigma_{i=1}^{N} \omega(r_i) [G_{Obs}(r_i)]^2} \right\}$ for PDF data was scaled by the number of independent parameters minus the number of free fitting parameters [37].) In Figure 5(a), we show the temperature-dependent Rw parameters for orthorhombic models ($P2_12_12_1$, $Pna2_1$, and $Pnma$) compared to the high-symmetry tetragonal ($P4/mbm$ and $I4/m$) and cubic ($Im\overline{3}$ and $Pm\overline{3}m$) models. In this PDF analysis, the orthorhombic space groups serve as simple models of the low-temperature structure to assess temperature-dependent trends in structure.

Examining Figure 5, it is seen that the $Pm\overline{3}m$ cubic structure does not represent the local structure for the entire temperature range measured. With reduced temperature, the best model is the $I4/m$ tetragonal structure between 350 K and 500 K followed by the $Pna2_1$ or $P2_12_12_1$ space groups for temperatures below $\sim 350$ K. Expanding the temperature range down to 10 K reveals multiple transitions if the Br and Cs atomic displacement parameters (ADPs) are examined. In Figure 5(b), we show the Br and Cs ADPs, $U_{Br}$ and $U_{Cs}$. Examination of the Br ADPs (for Br1, Br2, and Br3 sites) reveals a continuous transition near 170 K, a transition which onset near 300 K leading to an abrupt change near 350 K and a kink near 410 K. The transitions near 300 K and 170 K are particularly clear in the Cs and Pb ADPs shown in Figure 5(c). The 170 K transition is seen in the Raman data [see Figure 1(c)] as well as in these PDF results. No space group change is indicated for the transitions near 170 K. Hence, this is consistent with an order-disorder type continuous transition. We also note that in the single-crystal diffraction data discussed above, no new peaks are seen when comparing reciprocal space images at 200 K and 120 K. Figure 5(d) gives the full PDF derived temperature-dependent lattice parameters between 10 K and 500 K. Panels 5(b) and 5(c) reveal broad regions of structural change are seen in ADPs between 265 K and 345 K for the Br sites and between 170 K and 300 K for the Cs sites. The lower temperature limit in the Cs case is due to its higher mobility in the lattice compared to Br. This observation is consistent with the observed formation of twinning domains in the diffraction measurement shown in Figure 2(c). In contrast to the continuous changes observed in bulk lattice parameters derived from single-crystal diffraction measurement for all phase transition, the Cs, Pb, and Br ADPs, and PDF derived $a$ and $b$ lattice parameters change abruptly at $\sim 350$ K. In the bulk structural measurement, the atomic pairs are averaged over the whole volume of the measured single crystal. Since The wavelength ($\lambda = 0.41328$ Å) used in our single-crystal diffraction measurements gives an attenuation length of $\sim 110$ $\mu m$ which is significantly larger than the crystal dimensions. While local structural probes can resolve the local distortions into distinct atomic pairs. This observation in local structure indicates an abrupt change of domain structure revealing the first-order nature of the phase transition at $\sim 350$ K. It also supports the existence of local polar fluctuations reported previously [23]. We suggest here that such fluctuations may drive the isostructural first-order transition at 350 K. Significant deviation in Rw between the tetragonal and orthorhombic models occurs below 410 K indicating the existence of a local monoclinic phase for temperatures up to 410 K consistent with the single-crystal diffraction measurement results. These results are compatible with the recent nano-scale X-ray diffraction imaging results which found that the domains formed at 350 K are partially stable subject to cycling to lower and higher temperatures [38].

To explore specific atomic correlations in this material, X-ray absorption measurements were conducted between 20 K and room temperature. Three-component fits (Br-Pb, Br-Cs, and Br-Br pairs), over the R-space range $2.0 \le r \le 4.0$ Å, were made at each temperature between 20 K and 95 K, and a single-component fit (Br-Pb) was made for data above 95 K. The shells beyond the nearest neighbor Br-Pb shell (Br-Cs and Br-Br) are found to be suppressed for higher temperatures (above $\sim 100$ K, see Figure S16). The extracted widths for the Br-Pb shell were then fit to a simple Einstein model: $\sigma^2(T) = \sigma_0^2 + \frac{\hbar^2}{2\mu K_E \theta_E} coth(\frac{\theta_E}{2T})$ [39], where $\mu$ is the reduced mass for the atomic pair, and $\sigma_0^2$ represents the static disorder. This simple model represents the atomic pair motion as harmonic oscillations of a single effective frequency proportional to $\theta_E$. The fits yield an Einstein temperature of 104 K corresponding to an effective oscillator frequency of 72 $cm^{-1}$, indicating the extreme softness of the material. The low-temperature $\sigma^2$ values indicate static disorder for Br-Cs pairs, which is more than 20 times that for the first neighbor Br-Pb pairs and more than 2 times that for Br-Br pairs (Figure S16). Examination of the Br-Cs pair distribution derived from the single-crystal data at low temperature, reveals a broad spread in Br-Cs atomic distances consistent with this result (Figure S17). There is a large spread in the Br-Cs distribution in the $P2_12_12_1$ space group. This spread in positions becomes less broad at low temperatures (see Figure S18). Our molecular dynamics simulations reveal a larger dynamic change in Br-Pb and Br-Cs pair distributions in going from 100 K to 250 K, consistent with the proposed order-disorder transition (Figure S18). The Br-Cs correlations derived from these simulations show the most significant change with temperature consistent with the large thermal component to the disorder.

To understand the stable structural phases under strain, which may be present when films are grown on substrates with lattice mismatch, we conducted pressure-dependent structural measurements. Figure 6 exhibits the complementary high-pressure Raman and X-ray diffraction measurements conducted between ambient pressure and 17 GPa. Figure 6(b) shows the pressure dependence of specific Raman peaks revealing the onset and disappearance of features at the phase transitions. The sharpening of the peak near 70 cm$^{-1}$ in low-temperature ambient pressure measurements coincides with the appearance of the additional peak near 70 cm$^{-1}$ above 1 GPa. The very low pressure of the transition indicates the softness of the material. In previous pressure-dependent Raman measurements, the transition at $\sim 1$ GPa has been associated with Pb-Br distance length shrinkage and distortions of the PbBr$_6$ polyhedra [40]. Recall that the Raman peaks at 70 cm$^{-1}$ correspond to Cs atom shear-type motion triggered by asymmetric distortion modes of the PbBr$_6$ polyhedra. Hence, comparison of temperature-dependent and pressure-dependent Raman data suggests very low pressure recovers the low-temperature behavior of this system possibly with suppression of the disorder which onsets for temperatures above 170 K. It is expected that moderately straining these materials may lead to more stable phases with lower levels of electron scattering with better transport properties. High-pressure diffraction results are given in Figures 6(c) and 6(d). In Figure 6(d), an abrupt (first-order) transition is found near 1 GPa which is followed by a continuous transition that is completed near 2 GPa. Continuous transitions are also observed near 6 GPa and 13 GPa. Over the full range studied, we found five distinct structural phases (Phase I to V). Note that the transition near 1 GPa results in a splitting of peaks consistent with symmetry reduction below the ambient pressure space group.

To examine the behavior at low pressures, we expand the diffraction map between 0.6 and 3 GPa in Figure 7(a) and note that three phases are present. The region between 1 and 2 GPa reveals strong splitting of diffraction and Raman peaks. In the top panel of Figure 7(c), placing the sample on a glass slide yields Raman spectra characteristic of the normal bulk phase. Manually compressing the sample between a pair of glass slides and then remeasuring the spectra after removing the top slide produces a phase different from the original ambient pressure phase as shown in the lower panel. These additional features in the spectra are indicated as * symbols. By linearly extrapolating the high-pressure Phase II and Phase III frequencies back to the ambient pressure in Figure 7(b), we see that the new features are from a stabilized component of Phase II in the sample. Hence Phase II can exist at ambient pressure as a metastable phase after pressure release. This metastable form of Phase II is easy to achieve by releasing pressure rapidly after compression with glass slides. Measurements conducted two weeks after slide compression of the samples reveal the same spectra. On the other hand, after hydrostatic compression up to 17 GPa, slow release of pressure produces the original ambient pressure Raman and XRD patterns. Very low pressures can stabilize new phases and some of which will be stable under ambient conditions. These phases can be accessed when the material is prepared as a thin film since substrate-induced strain can readily place samples under these low pressures. We note that all observed transitions are reversible. Also, no amorphous phase is observed in the pressure range explored unlike the results on commercial polycrystalline materials examined in previous work [40].

The combined results of our experiments point to four distinct structural phases with cell dimensions $\sim 2a_p \times 2a_p \times 2a_p$ as temperature changes between 500 K and 10 K: a high-temperature cubic phase (above $\sim 410$ K), a high-temperature monoclinic phase (between $\sim 410$ K and $\sim 350$ K), a low-temperature monoclinic phase (between $\sim 350$ and $\sim 300$ K), and finally, a low-temperature monoclinic phase (below $\sim 300$ K). An isostructural phase transition is observed at $\sim 350$ K. An order-disorder transition is revealed at $\sim 170$ K by the loss of higher-order pair correlations in the X-ray absorption fine structure (XAFS) data and changes in the PDF ADPs. For all phase transitions, changes are evident in at least two of the independent measurements conducted with the $\sim 170$ K transition exclusively apparent in local structural studies. The phase transition near 300 K is found to be second-order and the other two higher-temperature phase transitions are found to be first-order in nature. The lower temperature transitions have not been identified in previous structural studies. The proximity of the second-order transition to room temperature suggests that room temperature measurements without temperature control may lead to uncertainty in the space group assignment (centrosymmetric vs. non-centrosymmetric). The local structural measurements indicate that the presence of distortions supports locally non-centrosymmetric symmetry for temperatures up to at least 500 K. The local structure is never in a cubic phase between 10 K and 500 K. In terms of the bulk structure, the phase below 300 K is non-centrosymmetric. The observed systematic absence violations and superlattice reflections reveal the inadequacy of the currently accepted room-temperature space group *Pnma*. We note that recent work on the $CsPbI_3$ system assumes the approximate space group assignments [41]. High dynamic-range synchrotron-based data sets should be collected to understand this general class of materials. Detailed studies such as those conducted here are needed to determine the correct space groups and corresponding physical properties. Such measurements will assist in the development of more accurate theoretical models of these important materials.

## 3 Conclusions

Understanding the basic physics underlying the properties of this material requires an accurate determination of the crystal structure. This is required to develop accurate potentials to predict the finite temperature properties such as transport and optical absorption. These critical properties are very sensitive to subtle structural details. New space group assignments are made for the temperature range from 100 K to 500 K. Structural measurements combined with second harmonic generation measurements reveal the absence of inversion symmetry below 300 K. In this temperature region, the true unit cell is a monoclinic $Pm$ cell. The space group is $P2_1/m$ between $\sim 300$ K and $\sim 410$ K and $Im\overline{3}$ above $\sim 410$ K. The structural parameters provided in this work will assist in the development of accurate models leading to the prediction of new and more efficient analogs of the all-inorganic $CsPbBr_3$ system. The first-order nature of the $\sim 410$ K and $\sim 350$ K transitions will impact transport properties if these materials are cooled rapidly once heated. Rapid cooling from above 350 K may freeze in the high-temperature phases. High-pressure experiments indicate multiple phase transitions at low pressure (near 1 and 2 GPa). These new phases will influence the properties of films grown on substrates with significantly different lattice constants from that of ambient $CsPbBr_3$. At low pressure, a phase is observed which can exist as a metastable phase at ambient pressure.

## 4 Experimental Methods and Modeling

*Differential Scanning Calorimetry*: Differential scanning calorimetry measurements were conducted under flowing $N_2$ gas using a Perkin Elmer DSC 6000. Measurements were made with a cooling/heating rate of 2 K/min.

*Raman Scattering Measurements*: Ambient pressure temperature-dependent Raman Spectra were measured with an excitation wavelength of 780 nm in backscattering geometry using a Thermo Scientific DXR Raman Microscope. A $50\times$ objective was used with a 15 mW laser power setting. The sample was found to be stable under this laser power after tests were done on a range of laser power values (0.1 to 15 mW). These measurements were conducted at the NJIT York Center. Samples return to the original phase after heating up to the maximum temperature of 830 K used in the experiments. High-pressure Raman measurements were conducted at the National Synchrotron Light Source II (NSLS-II) beamline 22-IR-1. Measurements were conducted in a symmetric cylindrical diamond cell with (100) oriented diamonds. For all Raman measurements, no change in the spectra was observed over time at a given pressure. Each pressure data set is comprised of sixty 10-second scans.

*X-ray Single-crystal Diffraction Measurements*: Diffraction measurements were conducted on an as-grown cube-shaped single crystal of edge $\sim 50\ \mu m$ at the Advanced Photon Source (APS) beamline 15-ID-D at Argonne National Laboratory using a wavelength of 0.41328 Å (30 keV). The data were collected with a PILATUS 1M detector between 100 K and 450 K in steps of 10 K (data are for increasing temperature). The NSF ChemMatCARS beamline is an undulator beamline. An undulator source does not output a continuous x-ray spectrum but a sharply peaked spectrum centered at the set energy, which is 30 keV in this case. In addition, the beamline utilized a Si (111) double crystal monochromator. The Si (222) Bragg reflection is forbidden. More importantly, the beamline has a harmonic rejection mirror to suppress the photons with energies above 30 keV. Hence the combination of tuned undulator energy, the use of a Si (111) monochromator, and a harmonic rejection mirror make Bragg peaks due to the $\lambda/2$ (60 keV) contamination impossible.

*Second Harmonic Generation Measurements*: The reflected SHG intensity was recorded as a function of the azimuthal angle $\phi$ (Figure 4). The incident ultrafast light source was of 50 fs pulse duration and 200 kHz repetition rate, and was focused down to a 20 $\mu m$ diameter spot on the sample with a fluence of $\sim 1$ mJ/cm$^2$. The RA-SHG patterns remain the same on increase of the fluency to $\sim 2$ mJ/cm$^2$

*Phonon DOS and Molecular Dynamics Simulations*: To determine force constants and phonon DOS for CsPbBr$_3$, density functional calculations in the projector augmented wave approach were carried out. Full structural optimization was conducted for both lattice parameters and atomic positions. The ground-state structure was optimized so that forces on each atom were below $2 \times 10^{-5}$ eV/Å. The phonon density of states and phonon displacement modes were derived from the computed force constants. *Ab Initio Molecular Dynamics* (MD) simulations were also conducted. A $2 \times 2 \times 2$ orthorhombic supercell (based on the optimized structure obtained above with 160 atoms) was utilized. In the MD simulations, the system temperature was set at 100 K, 250 K, and 500 K utilizing the NVT ensemble. MD time steps of 1 fs were used, with $\sim 2500$ time step for each simulation.

*X-ray Absorption Fine Structure Measurements*: Br K-edge XAFS spectra were collected at APS beam-line 20-BM at Argonne National Laboratory on single crystals ($\sim 2$ mm $\times$ 3 mm$\times$ 0.5 mm) in fluorescence mode (20 K to 125 K). Measurements were done in fluorescence mode on powders at NSLS-II beamline 7-BM (120 K to 300 K). Data were corrected for self-absorption. Reduction of the X-ray absorption fine structure data was performed using standard procedures.

*Pair Distribution Function Measurements*: Two independent pair distribution function data sets [140 to 500 K (run 1) and 10 to 200 K (run 2)] were collected at NSLS-II beamline 28-ID-2(XPD) at Brookhaven National Laboratory using a wavelength $\lambda = 0.1877$ Å and $\lambda = 0.1872$ Å for run1 and run2, respectively. Measurements utilized Perkin Elmer Area detectors with a sample to detector distance of $\sim 200$ mm. Exact detector to sample distances were derived by fits to Ni powder calibration standards. The Ni standard was used to determine set-up specific parameters (Q$_{damp}$ and Q$_{broad}$), which were held during data modeling.

*High-Pressure Powder Diffraction Measurements*: High-pressure powder diffraction measurements were performed at APS beamline 13-ID-D (GSECARS) at Argonne National Laboratory. The beam size used was 2.3 $\mu m$ (Vertical) $\times$ 3.1 $\mu m$ (Horizontal) with a wavelength of 0.3344 Å. A PILATUS 1M detector was used to collect the diffraction images. The sample-detector distance was 207.00 mm. The sample-detector geometry was calibrated with a LaB$_6$ powder NIST standard. The measurements were conducted with a diamond cell with 400 $\mu m$ cutlets. A 200 $\mu m$ thick rhenium gasket pre-indented to 42 $\mu m$ with a 200 $\mu m$ hole was used as the sample chamber. Neon was used as the pressure transmitting medium. Ruby balls and gold balls were placed near the pressed powder samples. The gold compression curve was used for pressure calibration.

*Statistical Analysis*: All data presented represent the mean values of multiple collected scans. The errors given for measurements are based on standard deviation on these average values. The data given are representative of the crystalline $CsPbBr_3$ material used in the measurements. All Raman and high-pressure diffraction data are shown after background subtraction for better visualization. Specific details for all methods, software, and references used for data processing and analysis are given in the supplementary document.

**Supporting Information**

Supporting Information is available from the Wiley Online Library or the author. The supplementary document includes: (1) full details of the experiments; (2) expanded detailed experimental spectra for Raman, X-ray diffraction, PDF, and XAFS measurements; (3) details on the phonon and molecular dynamics calculations; (4) tables of the representative single-crystal structural data.

**Acknowledgments**

This work is supported by NSF Award DMR-1809931. This research used beamlines 7-BM, 22-IR-1, and 28-ID-2 of the National Synchrotron Light Source II, a U.S. Department of Energy (DOE) Office of Science User Facility operated for the DOE Office of Science by Brookhaven National Laboratory under Contract No. DE-SC0012704. Single-crystal X-ray diffraction measurements were performed at NSF's ChemMatCARS Sector 15 at the Advanced Photon Source, which is principally supported by the National Science Foundation/Department of Energy under Grant NSF/CHE-1346572. Low-temperature XAFS measurements were conducted at Advanced Photon Source beamline 20-BM. Use of the Advanced Photon Source was supported by the U.S. Department of Energy, Office of Science, Office of Basic Energy Sciences, under Contract No. DE-AC02-06CH11357. This research used resources of the National Energy Research Scientific Computing Center, a DOE Office of Science User Facility supported by the Office of Science of the U.S. Department of Energy under Contract No. DE-AC02-05CH11231. L. Zhao acknowledges support by NSF CAREER Grant No. DMR-174774 and AFOSR YIP Grant No. FA9550-21-1-0065. Y. Yan acknowledges support from NSF award 1851747 for perovskite materials synthesis. We are indebted to C. L. Dias of the NJIT Physics Department for constructive criticism and physical insight, which has significantly improved the manuscript.

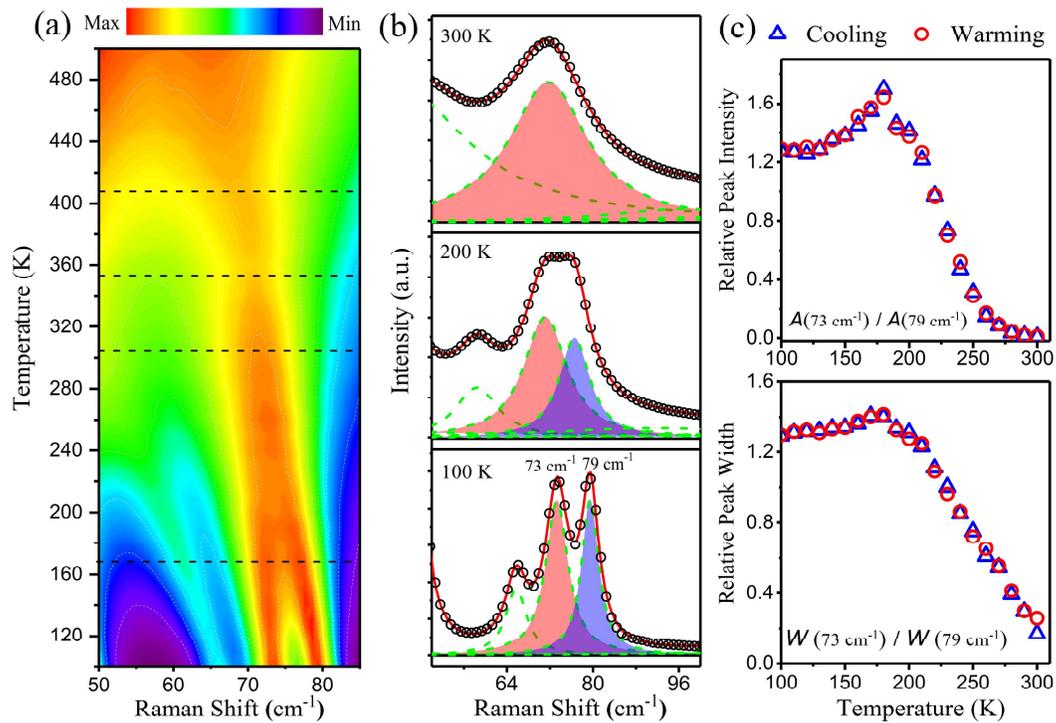

Figure 1: (a) Contour plot of the temperature dependence of the Raman spectra. (b) Low energy region of Raman spectra fitted by a sum of Lorentzian functions. The black open circles give the experimental data, the solid red line represents the sum of the fitting functions, and the dashed lines show the individual fitting components. A double-peak structure is shown, which is shaded in red and blue, for the modes at 73 cm$^{-1}$ and 79 cm$^{-1}$, respectively. The ratio of fitted peak area and peak width are shown in panel (c), indicating a phase transitions onset near 170 K and 300 K.

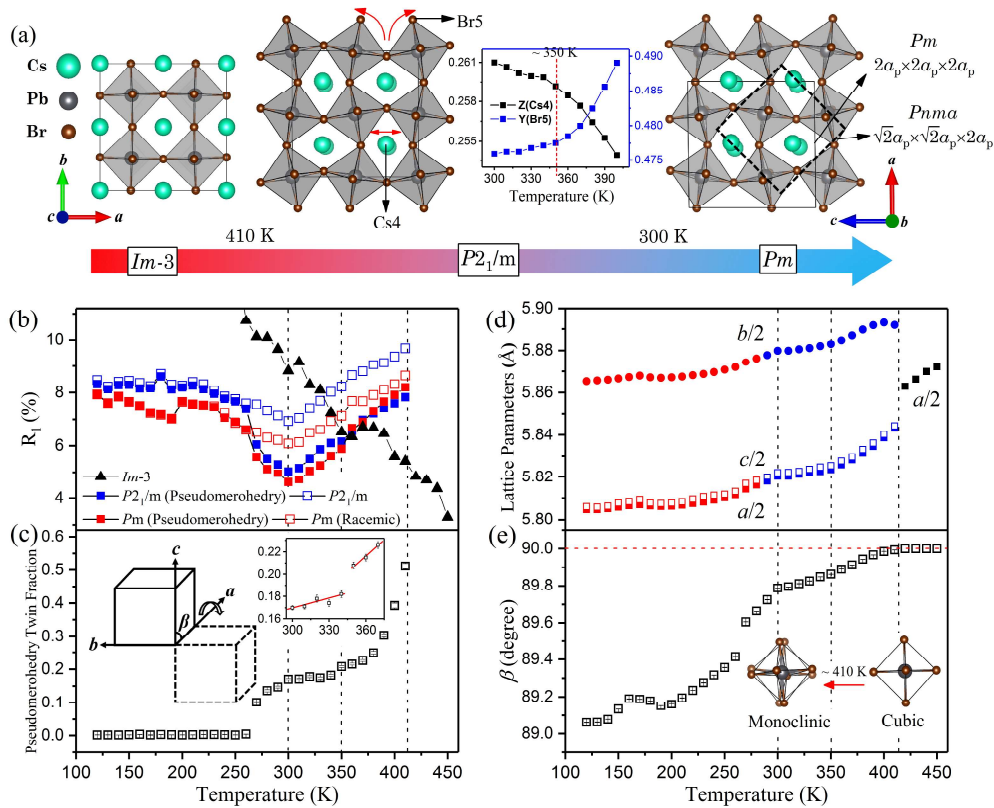

Figure 2: (a) The temperature-dependent structure of CsPbBr₃ from single-crystal synchrotron X-ray diffraction measurements. Above 410 K, the space group is cubic $Im\bar{3}$. Between 410 K and 300 K, the structure is monoclinic $P2_1/m$. The inset shows the Z fractional coordinate of Cs4 and the Y fractional coordinate of Br5 as a function of temperature. An isostructural phase transition is observed at 350 K. Below 300 K, the space group is $Pm$. The $Pm$ and previously reported $Pnma$ unit cells are given as the solid and dotted lines, respectively. (b) The quality-of-fit parameter, R1, of $Pm$, $P2_1/m$, and $I$-$3m$ structures. The close squares indicate the R1 parameters which incorporate a pseudomerohedry twin law matrix (1 0 0, 0 -1 0, 0 0 -1), while the open squares are the R1 parameters of racemic twinning for $Pm$ structure and no twinning for $P2_1/m$ structure. The temperature-dependent pseudomerohedry twin domain fraction is shown in (c). The single-crystal diffraction derived lattice parameters as a function of temperature are given in (d) and (e).

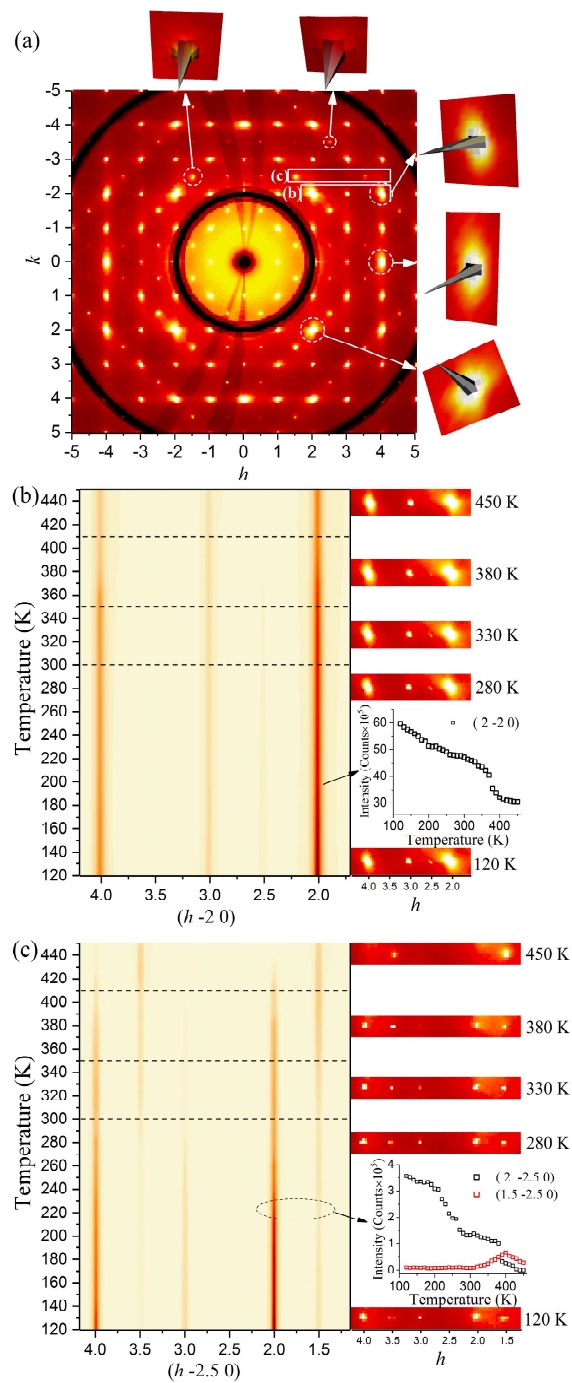

Figure 3: (a) Single-crystal X-ray diffraction reciprocal lattice image of the ($h\,k\,0$) plane at 450 K. The ($h\,k\,l$) grid corresponds to the previously reported $Pm\bar{3}m$ space group with lattice-constant $a_p \sim 5.87$ Å. The insets show the 3D intensity of some selected reflections with an asymmetric diffuse scattering background. Diffraction spots with half-integer $h$ and $k$ values are observed, indicating the correct lattice constant should be doubled. The temperature-dependent intensity of ($h$ -2 0) and ($h$ -2.5 0) reflections are given in panels (b) and (c), respectively, which are the selected regions shown in panel (a).

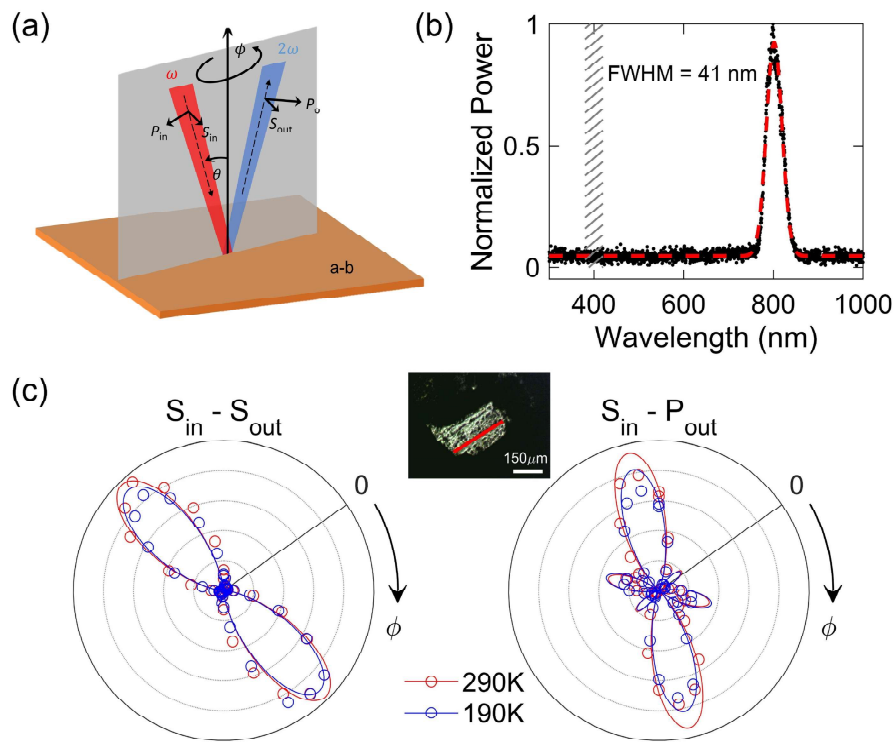

Figure 4: (a) Diagram of the RA-SHG setup. The fundamental beam is focused on the $ab$ plane of the crystal with a fixed small incident angle $\theta$. The scattering plane rotates about the surface normal by an angle $\phi$. (b) The spectrum of the incident fundamental light, centered at 800 nm with a FWHM of 41 nm. The shaded area at $400 \pm 20$ nm represents the bandpass region of the filter used to only collect SHG signals. (c) The RA-SHG patterns in $S_{in}$-$S_{out}$ and $S_{in}$-$P_{out}$ channels under 190 K (blue) and 290 K (red). $\phi = 0$ represents the direction of a natural edge of the sample. SHG barely shows any difference between data sets at these two temperatures. Inset: the picture of a measured $CsPbBr_3$ sample. The red bar marks its natural edge.

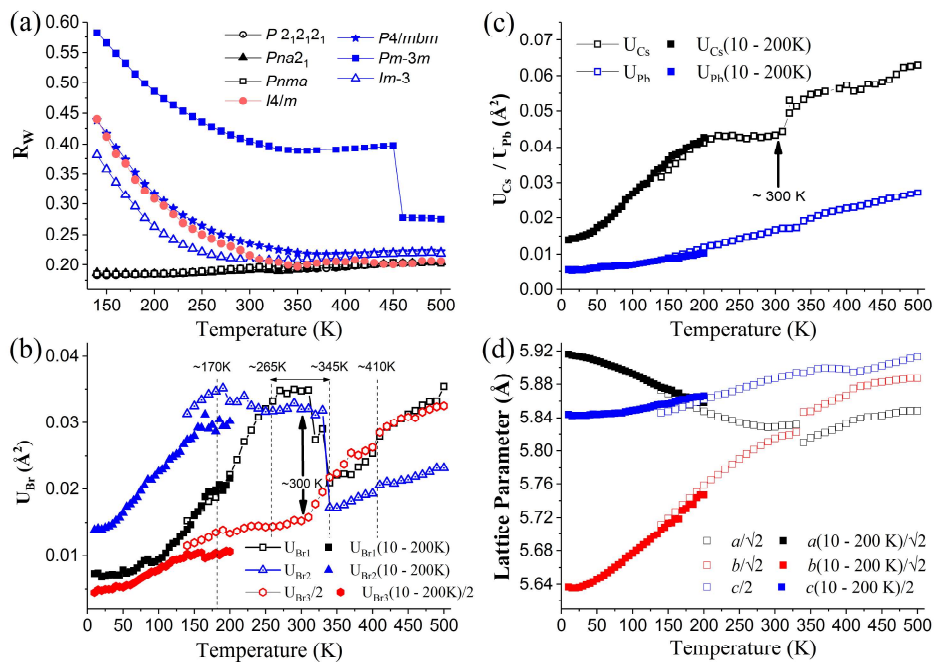

Figure 5: Results for local pair distribution function measurements. (a) The goodness-of-fit parameter, Rw, vs. temperature for different models, $P2_12_12_1$, $Pna2_1$, $Pnma$, $P4/mbm$, $Pm\bar{3}m$, $I4/m$ and $Im\bar{3}$. Atomic displacement parameters as a function of temperature for the Br, Cs, and Pb sites derived from the $Pna2_1$ model are shown in (b) and (c). In panels (b) and (c), solid symbols and open symbols are given for two independent data sets collected. The ADP parameters reveal clear structural changes at $\sim 170$ K and $\sim 410$ K. Broad regions of structural change are seen in ADPs between 265 K and 345 K for the Br sites and between 170 K and 300 K for the Cs sites (in panes (b) and (c)). (d) PDF derived temperature-dependent lattice parameters between 10 K and 500 K with an abrupt change near 350 K.

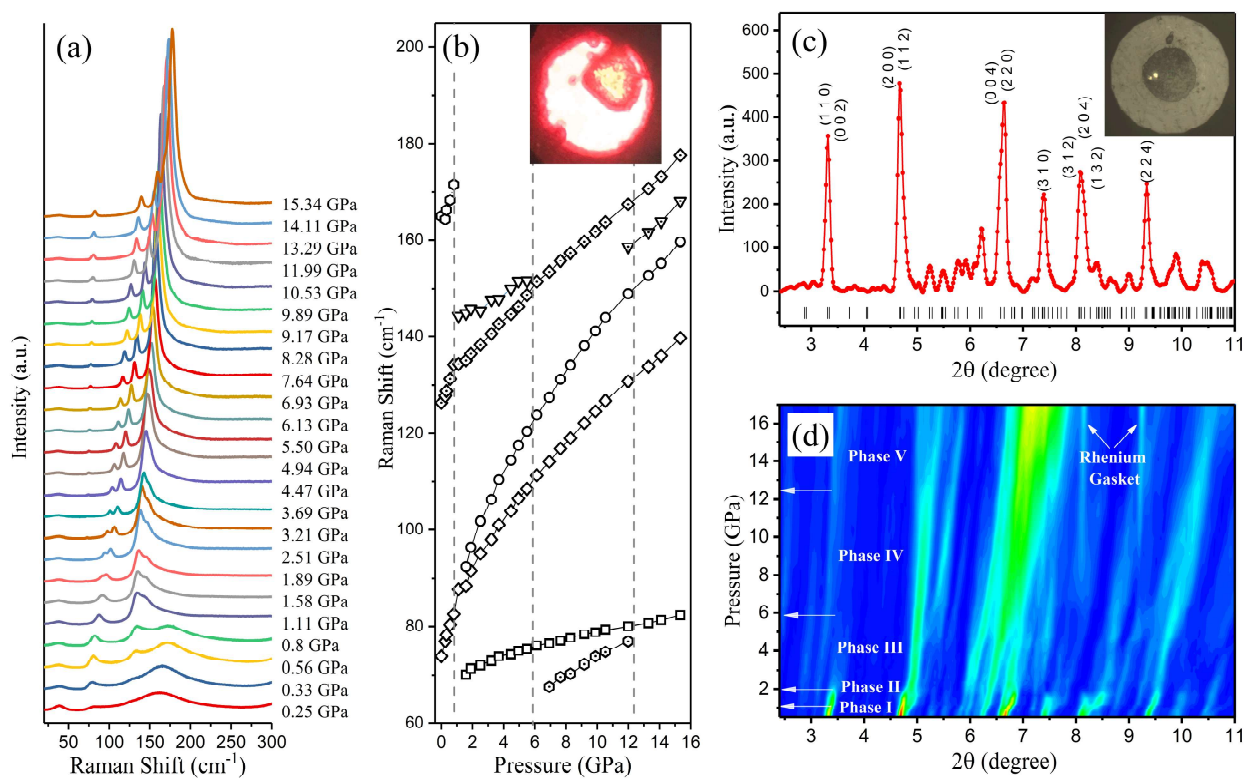

Figure 6: High-pressure structural changes at room temperature. (a) High-pressure Raman spectra for pressures between 0.6 and 15 GPa. The fitted Raman peak positions are shown in (b), indicating three phase transitions. The inset in (b) shows the sample in the diamond anvil cell with spot from a 646 nm laser. (c) Representative high-pressure powder X-ray diffraction pattern measured at 0.6 GPa. The inset shows the sample in the diamond anvil cell. (d) The 2D intensity plot of the pressure-dependent X-ray diffraction patterns indicates transitions at ∼ 1 GPa, 2 GPa, 6 GPa, and 13 GPa. The pressure-dependent structural phases are labeled Phase I to Phase V.

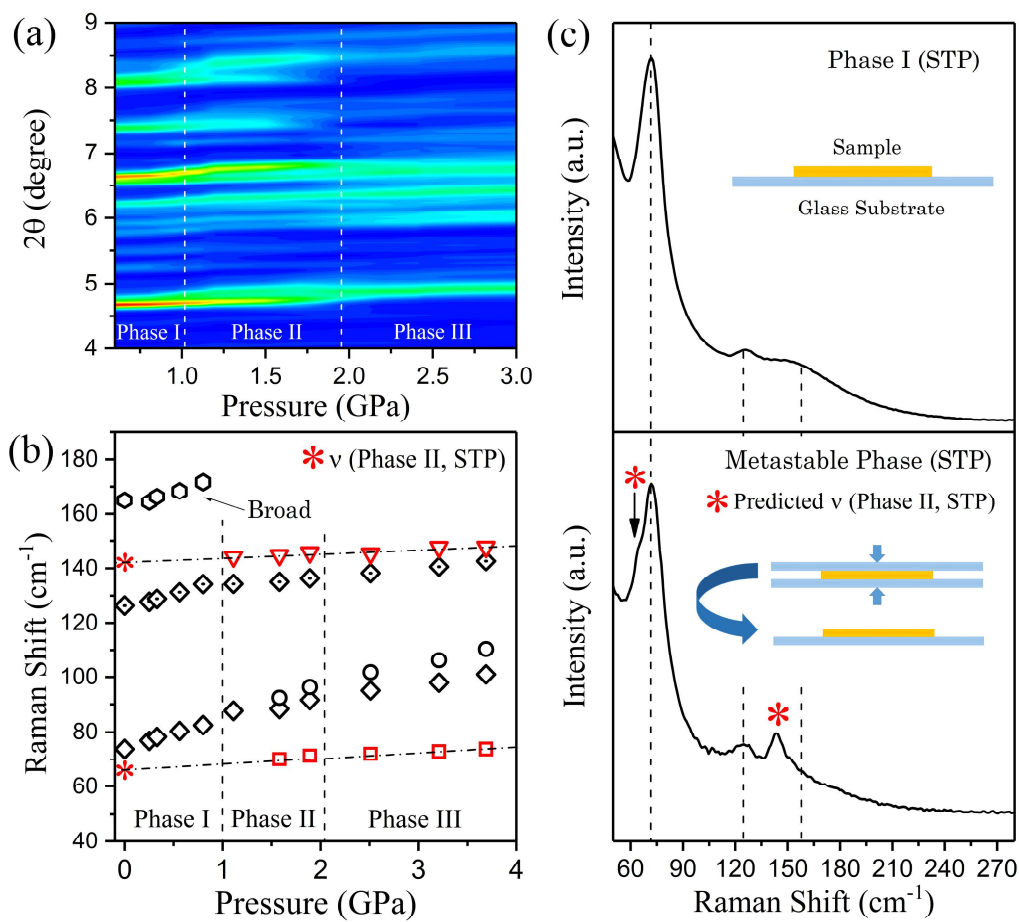

Figure 7: (a) Expanded region of the 2D X-ray diffraction map between 0.6 and 3 GPa showing a new phase that covers the region between 1 and 2 GPa. (b) The pressure-dependent Raman peak positions with dot-dashed lines giving predicted positions at standard temperature and pressure (STP) from linear extrapolation of the Phase II peaks, indicated with $*$ symbols. In the top panel of (c), the Raman spectrum of the original sample is shown while the spectrum of a sample after compression between glass slides is given in the bottom panel. The appearance of additional peaks is indicated by the $*$ symbols. Compression between the glass slides brings the samples into Phase II and a part of the sample is maintained in this phase after release of pressure (metastable Phase II).

# High-Resolution In-situ Synchrotron X-ray Studies of Inorganic Perovskite CsPbBr₃: New Symmetry Assignments and Structural Phase Transitions

S. Liu[1], A. R. DeFilippo[1], M. Balasubramanian[2], Z. Liu[3], S. G. Wang[4], Y-S. Chen[4], S. Chariton[4], V. Prakapenka[4], X. Luo[5], L. Zhao[5], J. San Martin[6], Y. Lin[6], Y. Yan[6], Y. S. Ghose[7], and T. A. Tyson[1]

[1]Department of Physics, New Jersey Institute of Technology, Newark, NJ 07102, USA
[2]Advanced Photon Source, Argonne National Laboratory, Argonne, IL 60439, USA
[3]Department of Physics, University of Illinois at Chicago, Illinois 60607-7059, USA
[4]Center for Advanced Radiation Sources, University of Chicago, Argonne, IL 60439, USA
[5]Department of Physics, University of Michigan, Ann Arbor, MI 48109, USA
[6]Department of Chemistry and Biochemistry, San Diego State University, San Diego, CA, 92182, USA
[7]National Synchrotron Light Source II, Brookhaven National Laboratory, Upton, NY 11973, USA

**(Supplementary Document)**

PbBr$_2$ (1.83 g, 5 mmol) and CH$_3$NH$_3$Br (560 mg, 5 mmol) were dissolved in 50 ml of dimethylformamide. The solution was heated slightly to obtain a transparent solution. This solution was further filtered through a compacted Celite column and the filtrate was collected. Two milliliters of this solution was transferred into an inner vial (5 ml in total vial volume) that was placed in a larger outer vial (25 ml in total volume) with 5 ml of toluene inside. Finally, the outer vial was carefully sealed. The diffusion of toluene from the outer vial into the inner vial was slow, and the crystallization process was maintained in a dark and undisturbed environment for at least three days. Orange block-shaped single crystals were obtained and characterized by X-ray diffraction. For powder sample derived experiments, crystals samples were crushed and sieved to obtain ~ 400 mesh powders. All measurements are based on crystal derived materials.

Differential scanning calorimetry measurements were conducted under flowing N$_2$ gas using a Perkin Elmer DSC 6000. Measurements were made at a cooling/heating rate of 2 K/min.

The rotational anisotropy second harmonic generation (RA-SHG) measurements were performed with the geometry shown in Fig.4(a) and spectrum of the incident beam shown in Fig. 4(b). The reflected SHG intensity was recorded as a function of the azimuthal angle $\phi$, while selecting either $S_{in}$-$S_{out}$ or $S_{in}$-$P_{out}$ polarization channels. In this experiment, the incident ultrafast light source was of 50 fs pulse duration and 200 kHz repetition rate, and was focused down to a 20 μm diameter spot on the sample and at a power of 0.7 mW, corresponding to a fluence of ~ 1 mJ/cm$^2$. The intensity of the reflected SHG was measured with a single photon counting detector.

Severe laser-induced lattice dynamics is not observed in our measurements on CsPbBr$_3$. RA-SHG patterns from the same sample at 290 K and the $S_{in}$-$S_{out}$ channel, taken under different optical fluence 1.1 and 2.3 mJ/cm$^2$, remain unchanged within the uncertainty level of the measurements. This indicates minimal lattice perturbation from the laser. Furthermore, the pulsed fs-laser source used in the RA-SHG experiments is of a 200 kHz repetition rate, corresponding to 5 μs separations between adjacent pulses. It is unlikely that the photoexcited lattice change/dynamics, if any, have a recovery timescale of microseconds making it detected by subsequent pulses.

Ambient pressure temperature-dependent Raman Spectra were measured with an excitation wavelength of 780 nm in backscattering geometry using a Thermo Scientific DXR Raman Microscope. A 50× objective was used with the laser power set at 15 mW. The sample was found to be stable under this laser power after tests were done on a range of laser power values (0.1 to 15 mW). Each temperature data set is comprised of one hundred 0.2-second scans. A Linkam Scientific THMS600 stage was used to measure the temperature-dependent Raman spectra. Only warming data are shown. Samples return to the original phase after heating up to the maximum temperature of 830 K used in the experiments. These measurements were conducted at the NJIT York Center.

High-pressure Raman measurements were conducted at the National Synchrotron Light Source II (NSLS II) beamline 22-IR-1 National. Measurements were conducted in a symmetric cylindrical diamond cell with (100) oriented diamonds. The diamond culet size was 500 μm, and tungsten gaskets were used. The pressure medium utilized was methanol:ethanol: water in a 16:3:1 ratio by volume. Pressure calibration was conducted using ruby fluorescence mainline shifts [1]. Pressure calibration measurements were made before and after each Raman spectrum was collected. In addition, calibration measurements as a function of position at multiple points (in the sample region of the gasket) at the highest pressure showed a high level of hydrostatic behavior of the pressure medium. Typical pressure errors are ± 0.10 GPa for pressures below ~ 4 GPa and ~ ±0.20 GPa for pressures above ~4 GPa. For this experiment, the custom micro-Raman system at beamline 22-IR-1 consisted of a 646 nm solid-state laser, a Princeton Instruments liquid-nitrogen cooled PyLoN CCD detector, a PI Acton SpectraPro SP-2556 Imaging Spectrograph, and a 20× objective. For all Raman measurements, no change in the spectra was observed over time at a given pressure. Each pressure data set is comprised of sixty 10-second scans.

Diffraction measurements on ~50 μm edge length crystals (cube-like shape) were conducted at the Advanced Photon Source (APS) beamline 15-ID-D (NSF's ChemMatCARS) at Argonne National Laboratory using a wavelength of 0.41328 Å (30 keV). The data were collected with a PILATUS 1M CdTe detector (by DECTRIS, maximum count rate = $10^7$ cps/pixel, counter depth =20 bit) between 100 K and 450 K in steps of 10 K (data are for increasing temperature). The data were processed using APEX3 (Bruker, 2016)[2]. The experimental reciprocal space precession images were generated using the same software. The simulated reciprocal space images were obtained using SingleCrystal 4.1.2 (CrystalMaker). The solution and refinement of the data were done using the program Olex2 [3] after the reflections were corrected for absorption using SADABS (with computed attenuation length = 105 μm). Anomalous scattering corrections were induced for all atoms. The values of f′ and f″ values for Br, Cs, and Pb at a wavelength of 0.41328 Å are 0.1889 and 0.9628, -1.7794 and 0.8050, and -1.4094 and 4.2152, respectively. Note that in the case of $CsPbBr_3$, the attenuation length of the x-rays beam with λ = 0.41328 Å (30.000 keV) is ~110 μm. This should be compared with standard Cu Kα (8.046 keV) or Mo Kα (17.48 keV) used in laboratory instruments, which yield attenuation lengths of ~10 μm and ~25 μm, respectively. The single-crystal goodness-of-fit parameters $R_1$ and $wR_2$ are defined as $R_1 = \sum ||F_0| - |F_c|| / \sum |F_0|$ and $wR_2 = \sum w(F_o^2 - F_c^2)^2) / \sum w(F_o^2)^2$, respectively. Detailed representative single-crystal solution results are presented in Tables S3 to S7.

We note that the NSF ChemMatCARS beamline 15-ID-D beamline was operated at 30 keV for these measurements. It is an undulator beamline. An undulator source does not output a continuous x-ray spectrum but a sharply peaked spectrum centered at the set energy, which is 30 keV in this case. In addition, the beamline utilized a Si (111) double crystal monochromator. The Si (222) Bragg reflection is forbidden. More importantly, the beamline has a harmonic rejection mirror to suppress the photons with energies above 30 keV. Hence the combination of tuned undulator energy, the use of a Si (111) monochromator, and a harmonic rejection mirror make Bragg peaks due to the λ/2 (60 keV) contamination impossible.

The final solutions presented are based on a comparison of the experimental reciprocal space images (see Figs. S6, S8, S13B below, for example) with the calculated images in addition to evaluation of the $R_1$ parameters. All observed Bragg peaks are accounted for in this approach (see Refs in [4] for a systematic approach to single crystal structure solution). Weak low-index reflections are essential to determining space groups [4(b)]. We have also accounted for twinning within the crystal [5].

To determine force constants and phonon DOS for CsPbBr$_3$, density functional calculations in the projector augmented wave approach were carried out utilizing the VASP code [6]. Full structural optimization was conducted for both lattice parameters and atomic positions. The LDA exchange functional (Ceperly and Alder as parameterized by Perdew and Zunger [7]) was used to obtain the relaxed structure. The ground-state structure was optimized so that forces on each atom were below 2 x $10^{-5}$ eV/Å. The optimized cell was found to be orthorhombic with volume = 8.3876 Å x 11.5197 Å x 7.5612 Å utilized ((4,4,4) gamma centered grid). Calculations for a 2×2×2 supercell with a gamma centered $k$-space grid were. The force constants were calculated in the frozen phonon approximation. The code Phonopy was utilized to determine the phonon density of states and phonon displacement modes from the force constants (Fig. S3, and Table S1) [8]. Gaussian broadening with full-width at half maximum of 7.1 cm$^{-1}$ was applied to each phonon DOS spectrum shown in Fig. S3(a).

To determine the low-temperature space group by modeling methods (DFT based on VASP), we initiated a structural optimization starting from the 120 K XRD Pm solution ($\sim 2a_P \times 2a_P \times 2a_P$ cell, ((4,4,4) gamma centered k-space grid)). In the first runs, the positions of the Pb atoms were fixed. The positions of all Br and Cs atoms were allowed to move, and the unit cell was free to adjust its shape to reduce the forces on all atoms to be less than 2.5 x $10^{-5}$ eV/Å. A second optimization cycle was conducted will all atoms and lattice parameters were free to adjust until the forces on atoms were minimized (again to less than 2.5 x $10^{-5}$ eV/Å). The structural optimization resulted in a monoclinic Pm cell with a = 11.293 Å, b= 11.518 Å, c=11.293 Å and b = 95.91°. We found the energy per CsPbBr3 for formula unit (f.u.) to be lower (E=-18.019 eV/f.u.) for this cell than that derived from fully optimizing a cell ((8,8,4) gamma centered k-space grid)) with dimension $\sim \sqrt{2}\ a_P \times \sqrt{2}\ a_P \times 2a_P$ (E=-17.944 eV/f.u.). These smaller cells result in a Pnma structure after optimization. Note that these calculations generate the zero temperature structure.

Molecular dynamics (MD) simulations were also conducted with the VASP code and projector-augmented wave (PAW) potentials [3]. The simulations were conducted as done in Ref. [9] for MAPbI$_3$ and used a 400 eV energy cutoff. A 2×2×2 orthorhombic supercell (based on the optimized structure obtained above with 160 atoms) was utilized. For separate MD simulations, the system temperature was set at 100, 250, and 500 K utilizing the NVT ensemble. MD time steps of 1 fs were used, with ~2500 time step for each simulation.

Br K-edge XAFS spectra were collected at APS beamline 20-BM at Argonne National Laboratory on single crystals (~2 mm x 3 mm x 0.5 mm) in fluorescence mode (20 K to 125 K). Higher temperature measurements were done in fluorescence mode with powders at beamline at NSLS-II beamline 7 BM (120 K to 300 K). Data were corrected for self-absorption. Reduction of the x-ray absorption fine-structure (XAFS) data was performed using standard procedures [10]. In the XAFS refinements, to treat the atomic distribution functions on equal footing, the Br K-edge spectra were modeled in R-space by optimizing the integral of the product of the radial distribution functions and theoretical spectra with respect to the measured spectra. Specifically, the experimental spectrum is modeled by,

$$\chi(k) = \int \chi_{th}(k, r) \, 4\pi r^2 g(r) dr$$ , where $\chi_{th}$ is the theoretical spectrum and g(r) is the real space radial

distribution function based on a sum of Gaussian functions ($\chi(k)$ is the measured spectrum) [11] at each temperature (as in Ref. [12]). For each shell fit, the coordination number (N) was held at the crystallographic value, but the position (R) and Gaussian width ($\sigma$) was fit to the data. the k-range $1.16 < k < 11.1$ Å$^{-1}$ and the R-range $1.96 < R < 4.00$ Å were utilized. Coordination numbers for the atomic shells were fixed to the crystallographic values. The Gaussian widths and positions were fit for each component

Two independent Pair distribution function (PDF) data sets (140 to 500 K (run 1) and 10 to 200 K (run 2)) were collected at NSLS-II beamline 28-ID-2 (XPD) beamline at Brookhaven National Laboratory using a wavelength $\lambda = 0.1877$ Å (run 1) and $\lambda = 0.1872$ (run 2). Measurements utilized Perkin Elmer Area detectors with a sample to detector distance of ~200 mm. Exact detector to sample distances were derived by fits to Ni powder calibration standards. The Ni standard was also used to determine set-up specific parameters ($Q_{damp}$ and $Q_{broad}$), which were held fixed for these samples. The range $Q_{mim} = 1.2$ Å$^{-1}$ and $Q_{max} = 24.5$ Å$^{-1}$ (run 1) was used in data reduction. (For run 2 the range was $Q_{mim} = 1.2$ Å$^{-1}$ and $Q_{max} = 22.5$ Å$^{-1}$ used.) All samples were measured in 1 mm Kapton capillaries with 50 micron thick walls. Scans were collected with blank capillaries to determine the background scattering. This background was subtracted from all datasets. The methods utilized for analysis of the PDF data are described in detail in Refs. [13]. For the fits in R-space, the range 2.0 < r < 30 Å was utilized. The time interval between temperature points was ~2 minutes. Combined with the small temperature steps, the approach kept the samples from being in a quenched state. For the PDF curves in Fig. 5(a), $R_W = \left\{ \frac{\sum_{i=1}^{N} w(r_i)[G_{Obs}(r_i) - G_{Calc}(r_i)]^2}{\sum_{i=1}^{N} w(r_i)[G_{Obs}(r_i)]^2} \right\}$, where $G_{Obs}$ and $G_{Calc}$ are the observed and calculated PDFs and $w$ is the weighting factor; $w(r_i) = 1/\sigma^2(r_i)$, where $\sigma$ is the estimated standard deviation on the data-point at position $r_i$, ref[10(b)]). Note that G(r) is the reduced atomic pair distribution function which oscillates about zero and is obtained directly from the scattering data, $S(Q)$, with $Q = \frac{4\pi \sin(\theta)}{\lambda}$. The function $G(r) = \frac{2}{\pi} \int_{Q_{min}}^{Q_{max}} Q(S(Q) - 1) \sin(Qr) \, dQ$ is related directly to the standard pair distribution function $g(r)$. $G(r) = 4\pi r \rho_0 (g(r) - 1)$ where $\rho_0$ is the number density of atoms. F(Q) in Fig. S14(b) is defined as $F(Q) = Q(S(Q) - 1)$. The PDF $G(r)$ includes all of Q-space between the limits of integration in Q-space and not just at the peak positions. Hence it captures the diffuse scattering [14].

High-pressure powder diffraction measurements were performed at APS beamline 13-ID-D (GESCARS) at Argonne National Laboratory. The beam size used was 2.3 μm (V) x 3.1 μm (H) with a wavelength of 0.3344 Å. A Pilatus 1M CdTe detector (by DECTRIS) was used to collect images. The sample-detector geometry was calibrated with a LaB$_6$ powder NIST standard. The sample-detector distance was 207.00 mm. The measurements were conducted with a diamond cell with 400 μm culets. A 200 μm thick rhenium gasket pre-indented to ~42 μm (with 200 μm hole) was used as the sample chamber. Neon was used as the pressure transmitting medium and Ruby balls and gold balls were placed near the pressed powder samples. Small pressure steps were enabled with the use of a gas membrane apparatus. At each pressure, 1-second exposures were conducted to acquire images. The sample was measured up to ~18 GPa and then released and remeasured. The ambient pattern was recovered on pressure release. Dioptas [15] were utilized to integrate the two-dimensional diffraction images (powder rings) to generate the intensity vs 2θ curves.

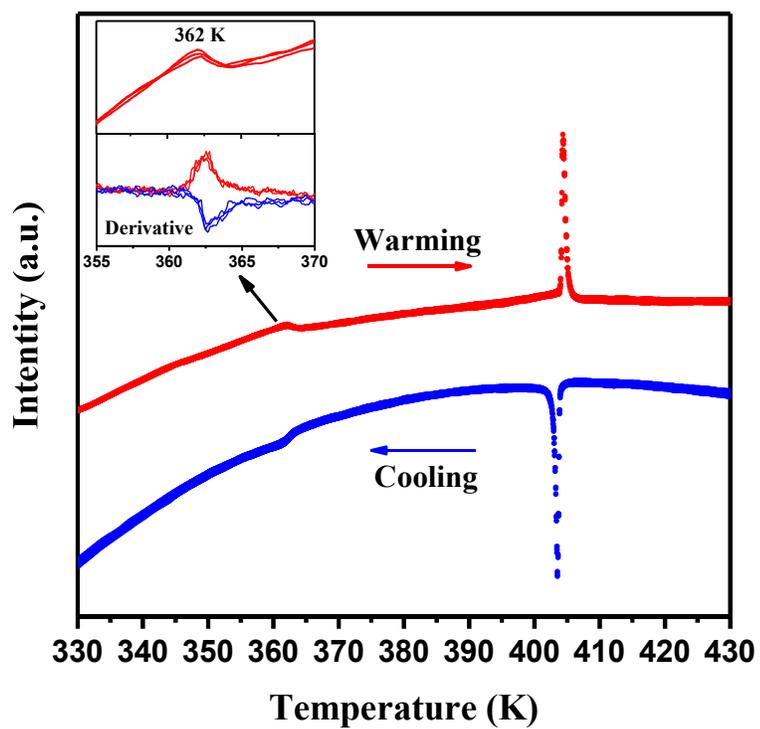

**Fig. S1.** DSC data showing first-order phase transitions at 362 K and 402 K in CsPbBr$_3$. The inset shows the derivative of the phase transition at 362 K. Both transitions reveal offsets on cooling and warming, pointing to their first-order nature. The inset displays multiple traces of the collected data.

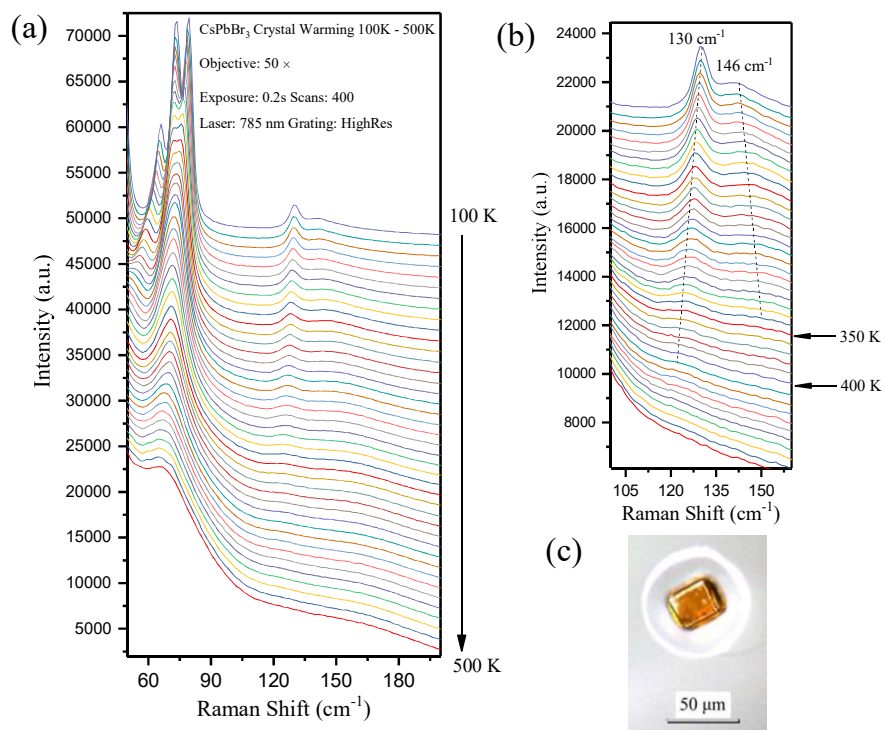

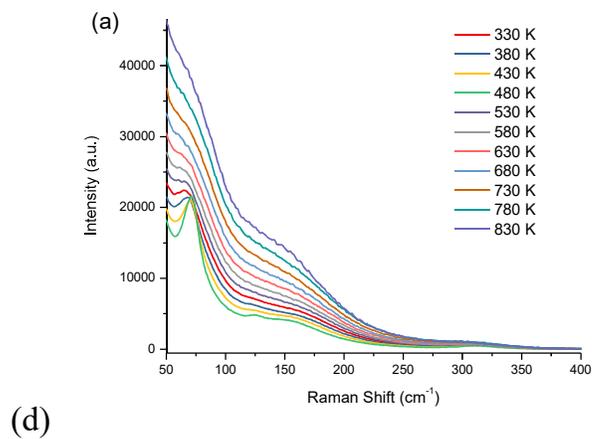

**Fig. S2.** (a) The Raman spectra of CsPbBr$_3$ from 100 K to 500 K. (b) Expansion near phonon modes at 130 cm$^{-1}$ and 146 cm$^{-1}$. (c) Representative single-crystal (in oil) used in single-crystal diffraction, Raman, and DSC measurements. (d) High-temperature Raman data between 330 and 830 K.

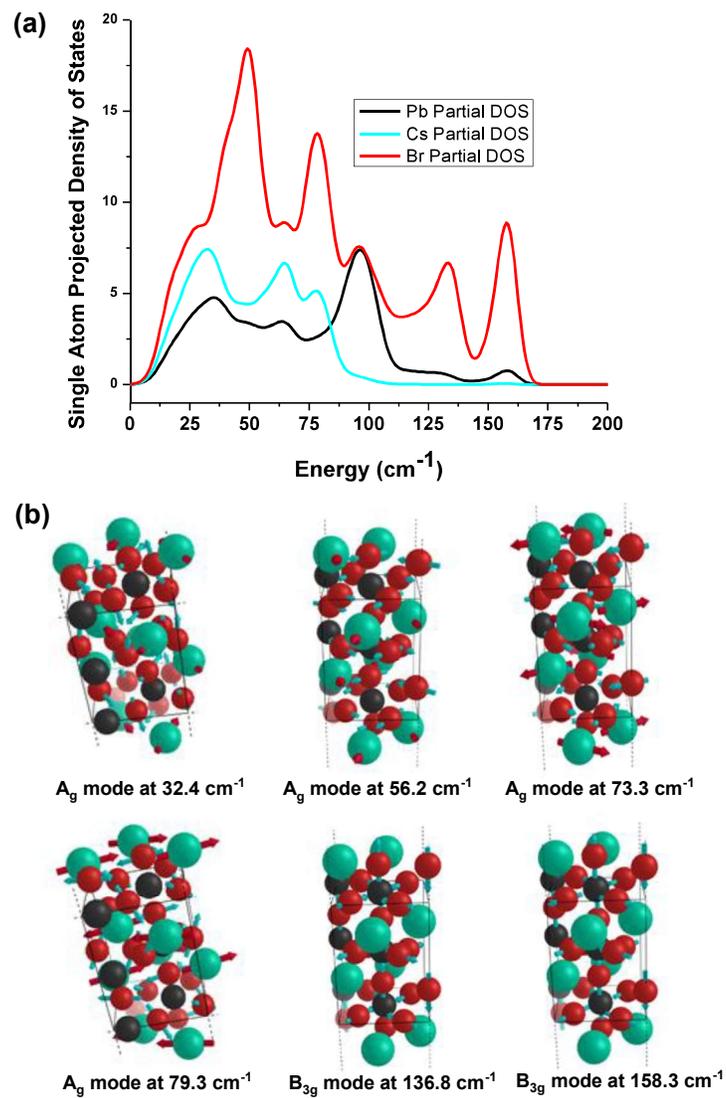

**(a)**

**(b)**

A$_g$ mode at 32.4 cm$^{-1}$    A$_g$ mode at 56.2 cm$^{-1}$    A$_g$ mode at 73.3 cm$^{-1}$

A$_g$ mode at 79.3 cm$^{-1}$    B$_{3g}$ mode at 136.8 cm$^{-1}$    B$_{3g}$ mode at 158.3 cm$^{-1}$

**Fig. S3**. (a) Partial phonon density of states derived from DFT simulations showing the Pb, Cs, and Br site projected components. (b) Selected Raman active phonon modes of CsPbBr$_3$ (see Table S1) indicating the motion of Cs (green), Br (red), and Pb (black) atoms.

**Table S1** Calculated Phonon Modes (Raman Modes Labeled)*

| Label | E (cm⁻¹) | Atomic Motion in Raman Active Mode |
|---|---|---|
| Au | 21.7 | |
| B1u | 25.5 | |
| B2u | 26.5 | |
| B3u | 29.3 | |
| Au | 30.2 | |
| Ag | 32.4 | Shear motion of layers containing Cs and Br |
| B2g | 34.7 | Shear motion of layers containing Cs and Br |
| Au | 35.5 | |
| Ag | 36.3 | Out of phase Breathing motion of Br shell about Cs |
| B1u | 36.5 | |
| B2u | 37.4 | |
| B3u | 38.3 | |
| Au | 39.1 | |
| B2g | 41.8 | Complex Cs and Br motion |
| Ag | 41.9 | Complex Cs and Br motion |
| B1g | 43.1 | Motion of subset of Br atoms only |
| B1u | 43.5 | |
| B3u | 46.0 | |
| B3g | 46.1 | complex Cs and Br motion |
| B2g | 47.6 | Complex Cs and Br motion |
| B2u | 48.4 | |
| B1g | 50.3 | Complex Cs and Br motion |
| B3u | 51.1 | |
| B3g | 51.9 | Complex motion of Cs and subset of Br |
| B1u | 52.3 | |
| Ag | 56.2 | Complex Cs and Br motion |
| B2g | 61.1 | Complex Cs and Br motion |
| Au | 63.1 | |
| B3u | 65.4 | |
| B1u | 65.6 | |
| B2u | 67.0 | |
| Ag | 73.3 | Complex Cs and Br motion |
| B1u | 75.4 | |
| B3u | 75.5 | |
| B1g | 75.6 | Complex Cs and Br motion |
| B2g | 75.8 | |
| Ag | 79.3 | Complex Cs and Br motion |
| B3g | 80.4 | Complex Cs and Br motion |
| B2g | 84.8 | Complex Cs and Br motion |
| B1u | 88.7 | |
| Au | 89.4 | |
| B2u | 91.8 | |
| B3u | 92.8 | |
| B3u | 93.5 | |
| B2u | 93.9 | |
| Au | 95.5 | |
| B2u | 96.7 | |
| B1u | 97.1 | |
| Au | 100.3 | |
| B1u | 100.4 | |
| B3u | 105.0 | |
| B1g | 134.3 | Out of phase Breathing mode of $PbBr_6$ Unit |
| Ag | 134.8 | Out of phase Breathing mode of $PbBr_6$ Unit |
| B3g | 136.8 | Out of phase Breathing mode of $PbBr_6$ Unit |
| B2g | 152.8 | In-phase breathing mode of $PbBr_6$ Unit |
| B1g | 153.4 | In-phase breathing mode of $PbBr_6$ Unit |
| B3g | 158.3 | In-phase breathing mode of $PbBr_6$ Unit |

*Raman Active Modes = 7Ag+ 5B1g +7B2g+5B3g

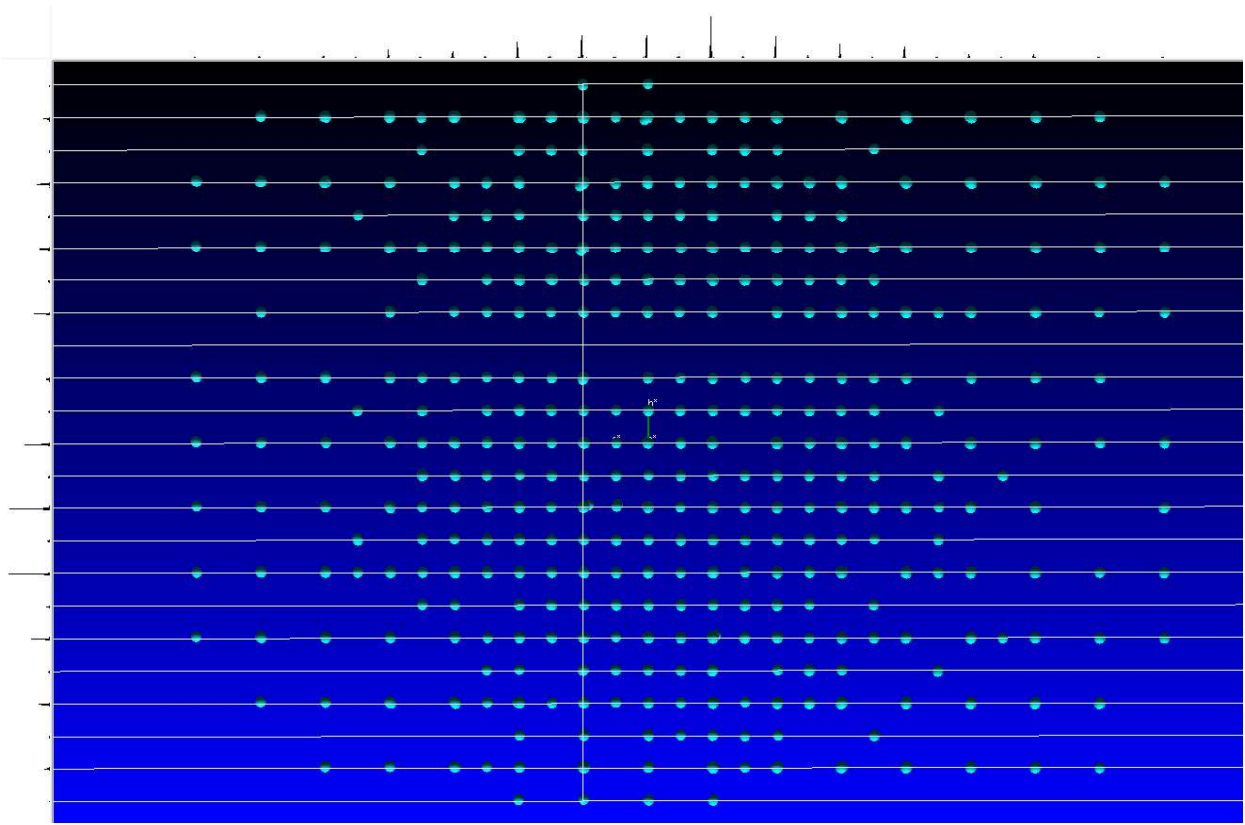

**Fig. S4.** View of reciprocal lattice points for data measured at 330 K.

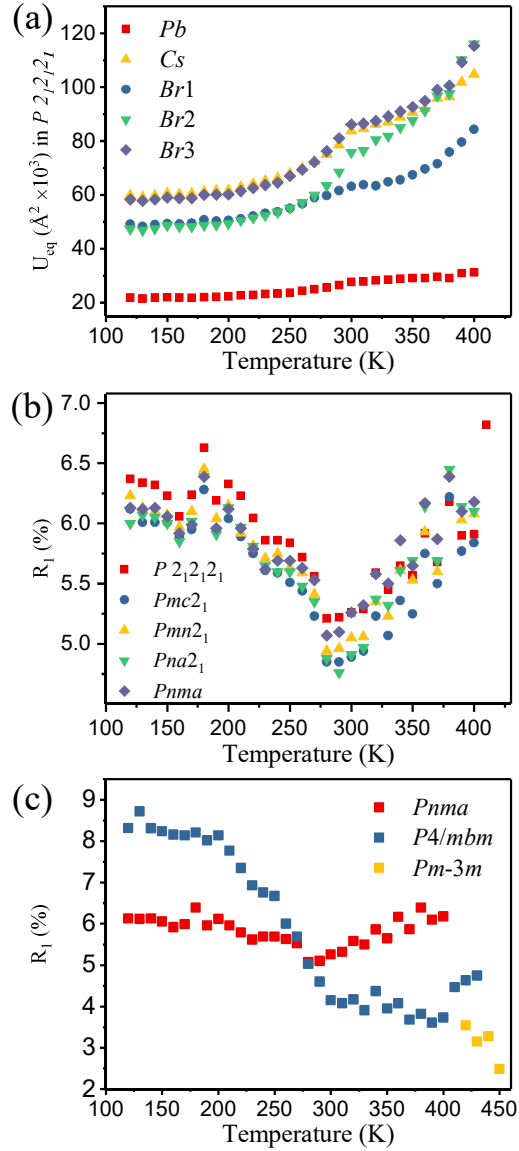

**Fig. S5.** (a) Temperature-dependent equivalent isotropic atomic displacement parameters ($Å^2 \times 10^3$) from single-crystal data for CsPbBr$_3$ in the $P2_12_12_1$ space group. $U_{eq}$ is defined as 1/3 of the trace of the orthogonalized $U_{IJ}$ tensor. (b) The R$_1$ parameters of all possible orthorhombic space group based on the cell dimension $\sim \sqrt{2} \, a_P$ x $\sqrt{2} \, a_P$ x $2 \, a_P$. Note that the R$_1$ values of the space groups are very close. However, systematic violations must also be examined (Table S2). (c) The R$_1$ based on the space groups in the literature. The unite cell for *Pnma* is $\sim \sqrt{2} \, a_P \times \sqrt{2} \, a_P \times 2 \, a_P$, *P4/mbm* is $\sim \sqrt{2} \, a_P \times \sqrt{2} \, a_P \times a_P$, and $\sim a_P \times a_P \times a_P$ for *Pm-3m* space group.

**Table S2-A.** Calculated Atomic Displacements of *Im*-3 structure at 450 K Compared to *Pm*-3*m* Structure

| *Pm*-3*m* Structure at 450 K | | | | Transform *Pm*-3*m* to *Im*-3 | | | | | *Im*-3 Structure at 450 K | | | | Atomic Displacements (Å) |
|---|---|---|---|---|---|---|---|---|---|---|---|---|---|
| | x | y | z | | x | y | z | | | x | y | z | |
| Pb1 | 0.5 | 0.5 | 0.5 | Pb1 | 0.75 | 0.25 | 0.25 | Compare → | Pb1 | 0.75 | 0.25 | 0.25 | 0.0000 |
| Cs1 | 0 | 0 | 0 | Cs1 | 0.5 | 0.5 | 0.5 | | Cs1 | 0.5 | 0.5 | 0.5 | 0.0000 |
| Br1 | 0 | 0.5 | 0.5 | Cs2 | 0.5 | 0 | 0.5 | | Cs2 | 0.5 | 0 | 0.5 | 0.0000 |
| | | | | Br1 | 0.5 | 0.25 | 0.25 | | Br1 | 0.5 | 0.24328 | 0.25652 | 0.1100 |

*Both structures were solved from the same data set.

**Table S2-B.** Calculated Atomic Displacements of *P*2$_1$/*m* structure at 380 K Compared to *P*4/*mbm* Structure

| *P*4/*mbm* Structure at 380 K | | | | Transform *P*4/*mbm* to *P*2$_1$/*m* | | | | | *P*2$_1$/*m* Structure at 380 K | | | | Atomic Displacements (Å) |
|---|---|---|---|---|---|---|---|---|---|---|---|---|---|
| | x | y | z | | x | y | z | | | x | y | z | |
| Pb1 | 0.5 | 0.5 | 0.5 | Pb1 | 0.5 | 0 | 0 | | Pb1 | 0.5 | 0 | 0 | 0.0000 |
| Cs1 | 1 | 0.5 | 0 | Pb2 | 0 | 0 | 0 | | Pb2 | 0 | 0 | 0 | 0.0000 |
| Br1 | 0.79195 | 0.70805 | 0.5 | Pb3 | 0 | 0 | 0.5 | Compare → | Pb3 | 0 | 0 | 0.5 | 0.0000 |
| Br2 | 0.5 | 0.5 | 0 | Pb4 | 0.5 | 0 | 0.5 | | Pb4 | 0.5 | 0 | 0.5 | 0.0000 |
| | | | | Cs1 | 0.75 | 0.25 | 0.25 | | Cs1 | 0.76111 | 0.25 | 0.24104 | 0.1666 |
| | | | | Cs2 | 0.75 | 0.25 | 0.75 | | Cs2 | 0.75828 | 0.25 | 0.73748 | 0.1752 |
| | | | | Cs3 | 0.25 | 0.25 | 0.75 | | Cs3 | 0.26259 | 0.25 | 0.74357 | 0.1650 |
| | | | | Cs4 | 0.25 | 0.25 | 0.25 | | Cs4 | 0.25724 | 0.25 | 0.23861 | 0.1575 |
| | | | | Br1 | 0.45805 | 0 | 0.25 | | Br1 | 0.4567 | -0.01829 | 0.24972 | 0.2160 |
| | | | | Br2 | -0.04195 | 0 | 0.75 | | Br2 | -0.04335 | -0.01855 | 0.74967 | 0.2191 |
| | | | | Br3 | 0.75 | 0 | 0.45805 | | Br3 | 0.74946 | 0.0188 | 0.45626 | 0.2225 |
| | | | | Br4 | 0.5 | 0.25 | 0.5 | | Br4 | 0.47942 | 0.25 | 0.48303 | 0.3110 |
| | | | | Br5 | 0.75 | 0 | 0.04195 | | Br5 | 0.75044 | -0.01753 | 0.04391 | 0.2078 |
| | | | | Br6 | 0 | 0.25 | 0 | | Br6 | -0.02017 | 0.25 | -0.01684 | 0.3063 |
| | | | | Br7 | 0.5 | 0.25 | 0 | | Br7 | 0.51575 | 0.25 | 0.01976 | 0.2946 |
| | | | | Br8 | 0 | 0.25 | 0.5 | | Br8 | 0.01827 | 0.25 | 0.52039 | 0.3192 |

*Both structures were solved from the same data set.

**Table S2-C.** Calculated Atomic Displacements of *Pm* structure at 250 K Compared to *Pnma* Structure

*Pnma* Structure at 250 K

| | x | y | z |
|---|---|---|---|
| Pb1 | 0 | 0.5 | 0.5 |
| Cs1 | 0.53865 | 0.25 | 0.49024 |
| Br1 | 0.2956 | 0.47265 | 0.2966 |
| Br2 | -0.00421 | 0.25 | 0.55377 |

Transform *Pnma* to *Pm*

| | x | y | z |
|---|---|---|---|
| Pb1 | 0.25 | 0.25 | 0.75 |
| Pb2 | 0.75 | 0.25 | 0.75 |
| Pb3 | 0.75 | 0.25 | 0.25 |
| Pb4 | 0.25 | 0.25 | 0.25 |
| Cs1 | 0.97579 | 0.0000 | 0.48555 |
| Cs2 | 0.47580 | 0.0000 | 0.98556 |
| Cs3 | 0.98555 | 0.0000 | 0.97579 |
| Cs4 | 0.48556 | 0.0000 | 0.47579 |
| Cs5 | 0.51445 | 0.5000 | 0.52420 |
| Cs6 | 0.01444 | 0.5000 | 0.02421 |
| Cs7 | 0.52421 | 0.5000 | 0.01445 |
| Cs8 | 0.02421 | 0.5000 | 0.51444 |
| Br1 | 0.29610 | 0.27735 | 0.49950 |
| Br2 | 0.22101 | 0.50000 | 0.77478 |
| Br3 | 0.72101 | 0.50000 | 0.27478 |
| Br4 | 0.79610 | 0.27735 | -0.00050 |
| Br5 | -0.00050 | 0.27735 | 0.29610 |
| Br6 | 0.72522 | 0.00000 | 0.77899 |
| Br7 | 0.70390 | 0.22265 | 0.20390 |
| Br8 | 0.70390 | 0.22265 | 0.50050 |
| Br9 | 0.20390 | 0.22265 | 0.00050 |
| Br10 | 0.49950 | 0.27735 | 0.79610 |
| Br11 | 0.22522 | 0.00000 | 0.27899 |
| Br12 | 0.00050 | 0.22265 | 0.70390 |
| Br13 | 0.27478 | 0.50000 | 0.22101 |
| Br14 | 0.77478 | 0.50000 | 0.72101 |
| Br15 | 0.27899 | 0.00000 | 0.72522 |
| Br16 | 0.77899 | 0.00000 | 0.22522 |

Compare →

*Pm* Structure at 250 K

| | x | y | z | Atomic Displacements (Å) |
|---|---|---|---|---|
| Pb1 | 0.25399 | 0.25016 | 0.75428 | 0.0112 |
| Pb2 | 0.75409 | 0.25013 | 0.75424 | 0.0119 |
| Pb3 | 0.75403 | 0.25019 | 0.25427 | 0.0115 |
| Pb4 | 0.25413 | 0.25015 | 0.25428 | 0.0116 |
| Cs1 | 0.98282 | 0 | 0.4879 | 0.0500 |
| Cs2 | 0.48154 | 0 | 0.98541 | 0.0661 |
| Cs3 | 0.98882 | 0 | 0.97364 | 0.0858 |
| Cs4 | 0.48949 | 0 | 0.47423 | 0.0787 |
| Cs5 | 0.51748 | 0.5 | 0.51899 | 0.1216 |
| Cs6 | 0.01793 | 0.5 | 0.0239 | 0.0643 |
| Cs7 | 0.52909 | 0.5 | 0.01639 | 0.0399 |
| Cs8 | 0.02872 | 0.5 | 0.51619 | 0.0410 |
| Br1 | 0.29878 | 0.27653 | 0.50528 | 0.0173 |
| Br2 | 0.22567 | 0.5 | 0.78773 | 0.0906 |
| Br3 | 0.72741 | 0.5 | 0.28772 | 0.0951 |
| Br4 | 0.79892 | 0.27659 | 0.00504 | 0.0148 |
| Br5 | 0.00244 | 0.27735 | 0.30208 | 0.0132 |
| Br6 | 0.73088 | 0 | 0.78717 | 0.0410 |
| Br7 | 0.50355 | 0.22181 | 0.21042 | 0.0200 |
| Br8 | 0.70625 | 0.22159 | 0.50595 | 0.0210 |
| Br9 | 0.20649 | 0.22186 | 0.00621 | 0.0177 |
| Br10 | 0.50246 | 0.27802 | 0.80184 | 0.0138 |
| Br11 | 0.23105 | 0 | 0.28739 | 0.0442 |
| Br12 | 0.00334 | 0.22343 | 0.70941 | 0.0147 |
| Br13 | 0.27915 | 0.5 | 0.23105 | 0.0566 |
| Br14 | 0.77877 | 0.5 | 0.73025 | 0.0469 |
| Br15 | 0.28251 | 0 | 0.73641 | 0.0695 |
| Br16 | 0.78335 | 0 | 0.23748 | 0.0823 |

*Both structures were solved from the same data set.

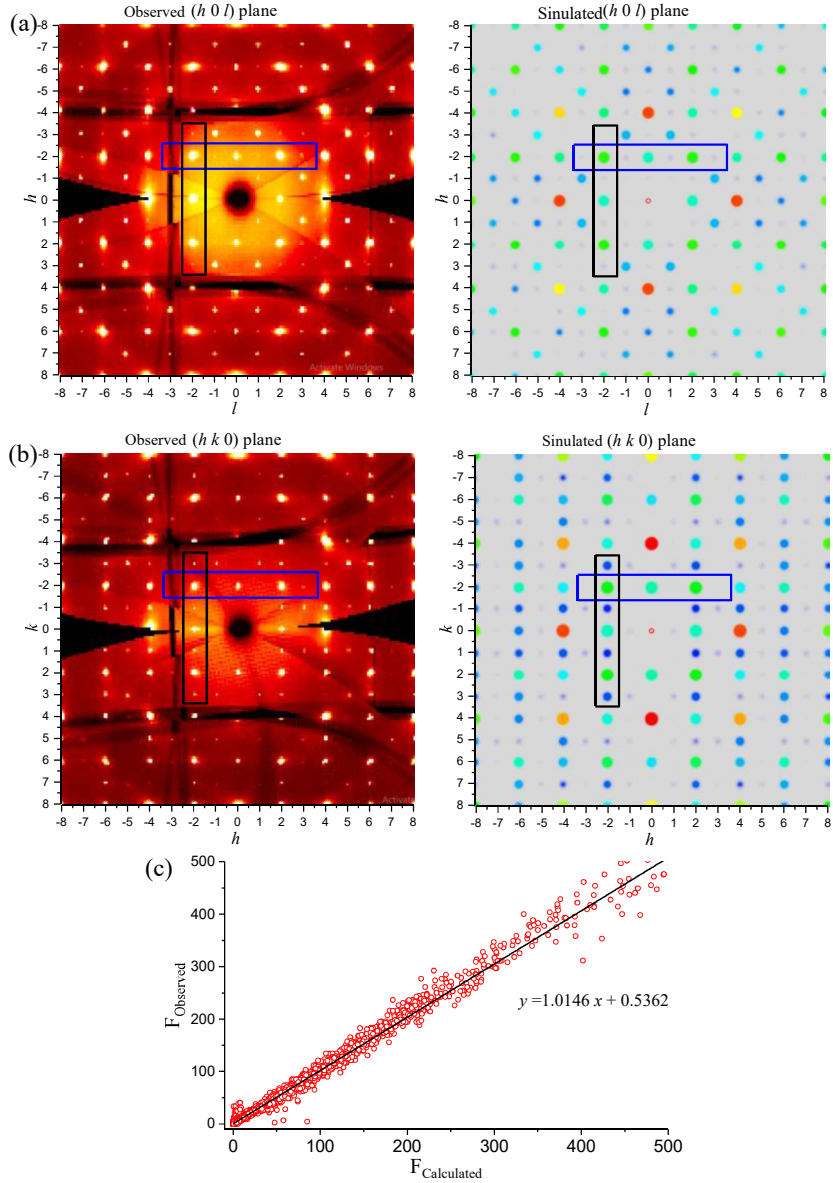

**Fig. S6.** Experimental reciprocal space precession images at 380 K for the (a) ($h$ 0 $l$) plane and (b) ($h$ $k$ 0) plane indexed with a $2a_P \times 2a_P \times 2a_P$ cell compared to the simulated pattern using the $P2_1/m$ crystal structure solution ($2a_P \times 2a_P \times 2a_P$). The $b$ axis is the long axis. Observe the qualitative similarity in the weak peaks representing the doubled cell relative to the simple cubic cell ($a_P \times a_P \times a_P$). Rectangles with the same color cover equivalent regions in measured and simulated images. In the calculation, the high-intensity peaks are in the red region of the color spectrum and the low-intensity peaks are on the blue end of the spectrum. The size of the peaks shown also indicates their intensities. (c) The $F_{Observed}$ vs $F_{Calculated}$ is fitted by a linear function at 380 K.

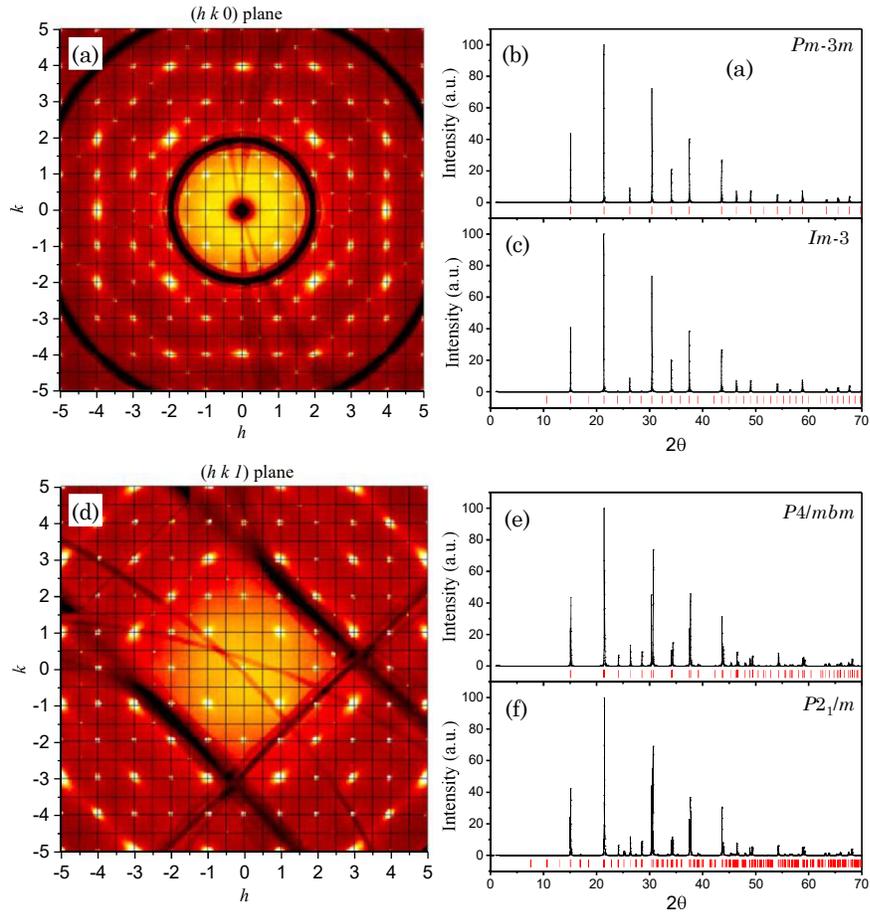

**Fig. S7.** (a) Single-crystal X-ray diffraction reciprocal space images of the ($h$ $k$ 0) planes of CsPbBr$_3$ at 450 K. The ($h$ $k$ $l$) grid corresponds to the previously reported *Pm*-3*m* space group with a lattice constant $a =$ 5.87 Å. Diffraction spots with half-integer values are observed, indicating that the correct lattice constant should be doubled. The simulated powder diffraction patterns of the *Pm*-3*m* and the newly proposed *Im*-3 space group based on the solved structure from single-crystal diffraction refinement are given in panels (b) and (c), respectively. (d) Single-crystal X-ray diffraction reciprocal space images of the ($h$ $k$ $l$) planes of CsPbBr$_3$ at 360 K. The grid corresponds to the previously reported *P*4/*mbm* space group with unit cell dimension: $\sqrt{2}\,a_P \times \sqrt{2}\,a_P \times a_P$. Note the presence of half-integer peaks. The simulated powder diffraction patterns of *P*4/*mbm* and the newly proposed *P*2$_1$/*m* space group based on the solution of the single-crystal diffraction refinement are given in panels (e) and (f), respectively. The wavelength for the simulated powder diffraction patterns is 1.54059 Å (Cu-Kα). Powder diffraction measurements are not adequate to distinguish between the *P*4/*mbm* and *P*2$_1$/*m* space groups in the 360 K data and between the *Pm*-3*m* and *Im*-3 space groups in the 450 K data. We also note that above 360 K, the *Im*-3*m* and *Im*-3 space groups have similar R$_1$ parameters. However, the Cs ADPs of *Im*-3*m* are highly anomalous (dramatically reduced in size with temperature increase). The anomalous behavior is due to the presence of distortions in CsPbBr$_3$ not supported by the high symmetry *Im*-3*m* space group.

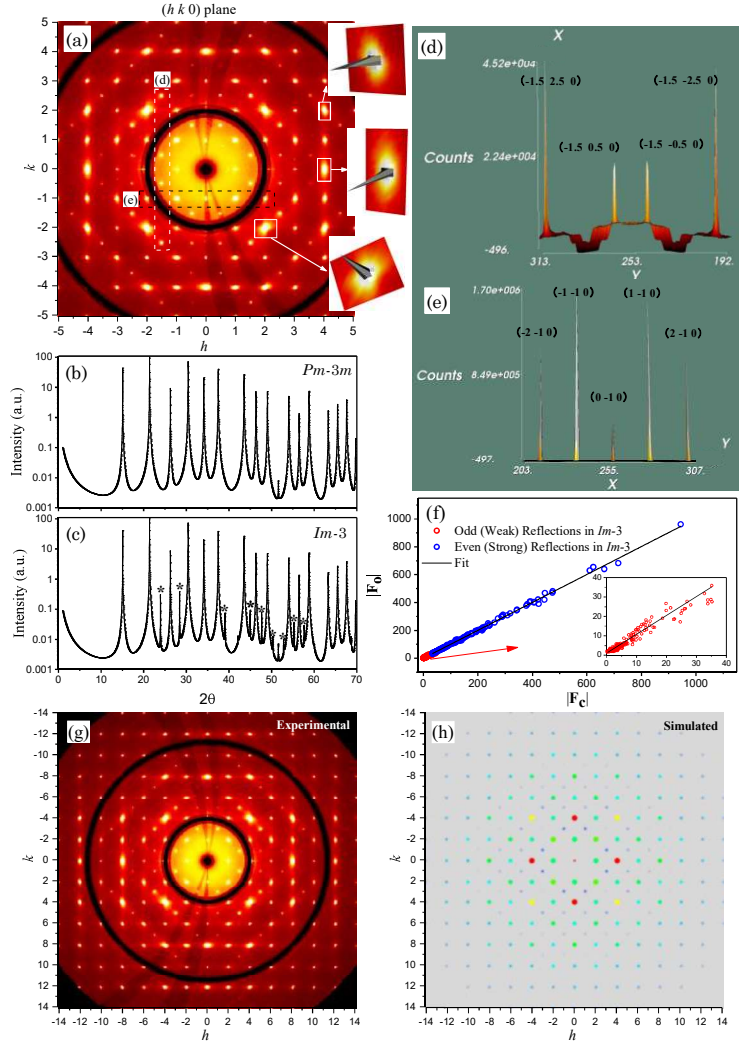

**Fig. S8.** (a) Single-crystal X-ray diffraction reciprocal space images of the (*h k* 0) planes of CsPbBr₃ at 450 K. The (*h k l*) grid corresponds to the *Pm-3m* (#221) space group with unit cell dimension: $\sim a_P \times a_P \times a_P$ where $a = 5.87$ Å. However, the presence of half-integer peaks indicates the *Pm-3m* simple cell is incorrect. The inset shows the 3D intensity of some selected reflections with an asymmetric diffuse scattering background. (b) Simulated *Pm-3m* powder XRD pattern with cell dimension: $\sim a_P \times a_P \times a_P$. (c) Simulated *Im-3* (#204) powder XRD pattern with cell dimension: $\sim 2a_P \times 2a_P \times 2a_P$. The intensity (y-axis) is displayed on a log scale. The additional features in the spectra are indicated with asterisks (*) symbols corresponds to the half-integer peaks in reciprocal lattice shown in panel (a). In the reciprocal space images, an examination of the intensities of half-integer reflections, (-1.5 *k* 0) is shown in panel (d), and the integer reflections, (*h* -1 0) is in panel (e), reveals the intensity of half-integer peaks are $\sim 10^2$ times weaker than the integer peak intensity but $\sim 10$ times stronger than the intensity of the background. Returning to the powder diffraction simulations [panel (c)], it is observed that the weak peaks (*) are of the same level relative to the main peaks as what is seen in the reciprocal space images (d) for the single-crystal measurements. Hence fitting powder data to the *Pm-3m* structure will not be strongly influenced by the exclusion of these additional peaks. Consequently, powder diffraction can not be used to ascertain the space group. A plot of |F_o| (observed) vs. |F_c| (calculated) parameters at 450 K for the strong (even-integer) and the weak satellite (odd-integer) reflections in *Im-3* space group with a linear fit (black line). The inset is the same plot with data for the odd-integer reflections. (g) Experimental reciprocal space image at 450 K for the (*h k* 0) plane compared with the simulated pattern (h) for the *Im-3* structure ($2a_P \times 2a_P \times 2a_P$). In the calculation, the high-intensity peaks are in the red region of the color spectrum and the low-intensity peaks are on the blue end of the spectrum. The size of the peaks shown also indicates their intensities.

**Table S3.** Refined Structural Parameters Utilizing the Weak Reflections Exclusively at 450 K in *Im*-3 Space Group

| Atoms | x | y | z | $U_{eq}$ (Å$^3$×10$^3$) |
|---|---|---|---|---|
| Pb1 | 7500 | 2500 | 2500 | 37(3) |
| Cs1 | 5000 | 5000 | 5000 | 75(5) |
| Cs2 | 5000 | 0 | 5000 | 75(5) |
| Br1 | 5000 | 2491(6) | 2509(6) | 81(4) |

Space Group: *Im*-3
$a$ = 11.7444 (3) Å, Dx = 4.755 g/cm$^3$
Measurement Temperature: 450 K
Crystal Dimensions (diameter): ~ 50 µm
Wavelength: 0.41328 Å
2θ range for data collection: 2.852° to 40.228°
Index ranges: -18 ≤ h ≤ 18, -18 ≤ k ≤ 18, -18 ≤ l ≤ 13
Reflections collected: 26205
Independent reflections: 513
Number of fitting parameters: 7
Largest diff. peak/hole: 0.38/-0.39 e/Å$^3$
$R_1$ = 50.18 %, w$R_2$ = 71.60 %, Goodness of Fit = 2.112

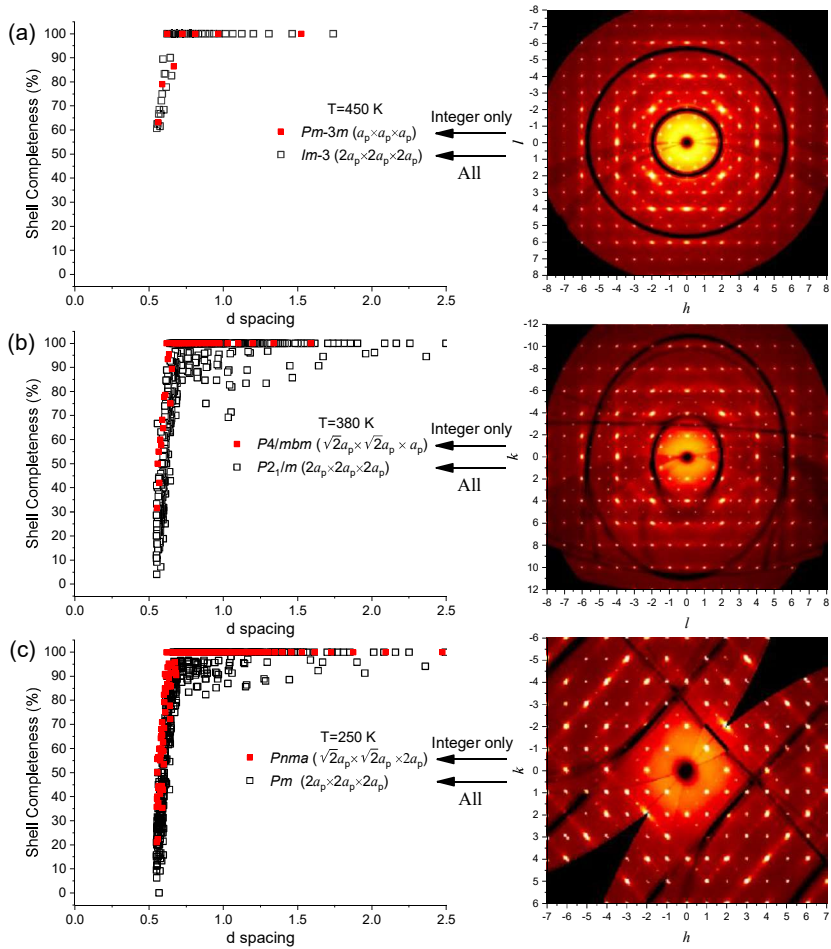

**Fig. S9.** The completeness as a function of d spacing is given for both the previously reported models and new models. Reflections are binned in d space using 20 bins for all models. This gives a representative sample of the density of reflections. Note that not all reflections are captured on this coarse grid. The reciprocal lattice precession images are also present to better clarify where are the additional reflections occur in the raw data. The ($h\ k\ l$) grids in these images are based on the old unit cell dimensions. The shell completeness of the old model is presented as the solid red square symbols and the new model as open square symbols. Note that, for the old models, mainly strong reflections (integer reflections) in the full data set are utilized. While for the new model, all the reflections (both integer and half-integer reflections) are captured and utilized in the structural refinement. The half-integer reflections are not fitted in the old models. In panel (c), it's not surprising that some weak reflections appear at low temperatures which are difficult to be captured completely. In our structural solutions, the overall completeness for the $P2_1/m$ and *Pm* solution is >97% while for *Im*-3 is >99%. The completeness of old models is >99% since they require only a subset (dominant reflections only) of the measured reflections.

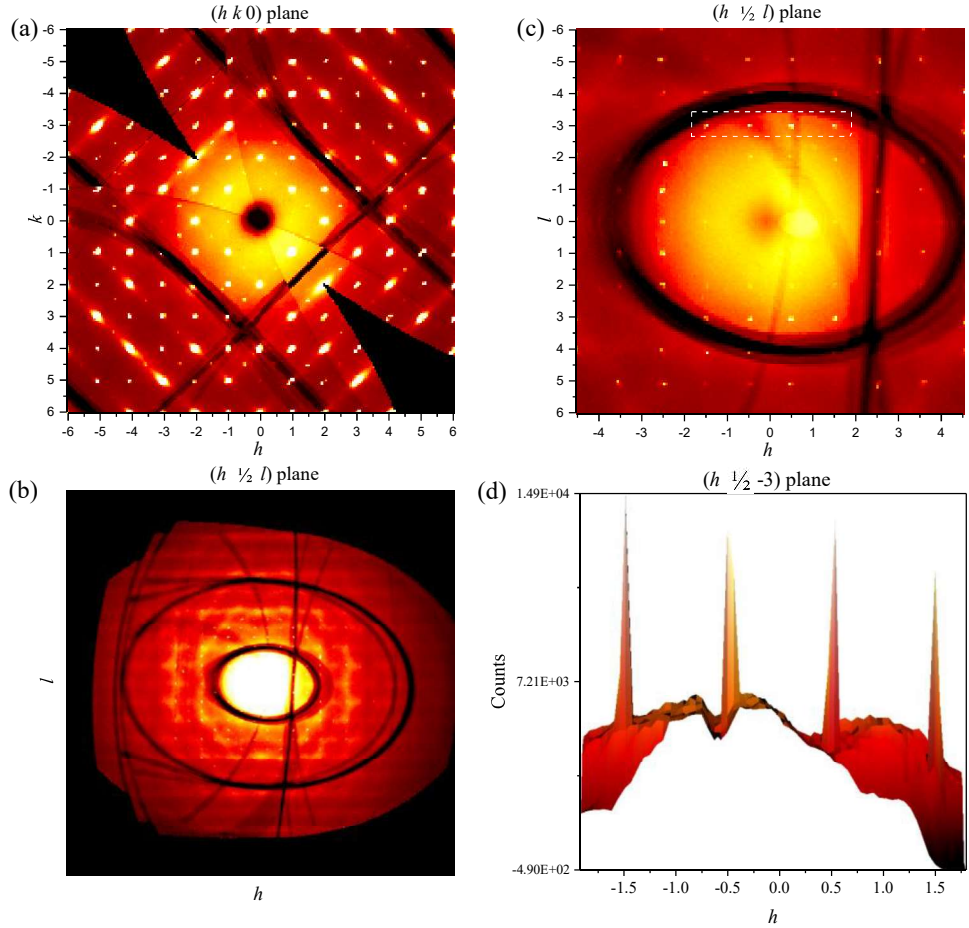

**Fig. S10.** (a) Single-crystal X-ray diffraction reciprocal space image of the ($h\ k\ 0$) plane at 200 K. The grid corresponds to the previously reported orthorhombic (*Pnma*) unit cell dimension $\sim \sqrt{2}\ a_P \times \sqrt{2}\ a_P \times 2a_P$. The half-integer reflections can be observed along the diagonal. (b) The reciprocal space image of the ($h\ \frac{1}{2}\ l$) plane shows the half-integer reflections, and the expanded image is given in (c). (d) The intensity map for the ($h\ \frac{1}{2}\ \text{-3}$) line corresponds to the selected region in panel (c). We found that these reflections are weak ($\sim 10$ times the background intensity). Careful considerations showed that all these previously unfitted weak reflections can indeed be indexed on a primitive monoclinic supercell with $a$ =11.6126(6) Å, $b$ =11.7344(6) Å, $c$ =11.6156(5) Å, and $\beta$ =89.1610(10). Considering all observed reflections, the space group can no longer be taken as *Pnma* space group. The true space group is *Pm* with unit cell dimensions $\sim 2a_P \times 2a_P \times 2a_P$.

**Table S4-A.** Number of Systematic Absence Violations*

| Temperature (K) | Space group | | | | |
|---|---|---|---|---|---|
| | $P2_12_12_1$ | $Pmc2_1$ | $Pmn2_1$ | $Pna2_1$ | $Pnma$ |
| 100 | 7 | 90 | 47 | 137 | 137 |
| 110 | 6 | 98 | 52 | 150 | 150 |
| 120 | 6 | 79 | 46 | 125 | 125 |
| 130 | 5 | 97 | 53 | 150 | 150 |
| 140 | 5 | 77 | 46 | 123 | 123 |
| 150 | 6 | 85 | 58 | 143 | 143 |
| 160 | 6 | 82 | 54 | 136 | 136 |
| 170 | 8 | 90 | 56 | 146 | 146 |
| 180 | 6 | 71 | 46 | 117 | 117 |
| 190 | 3 | 90 | 46 | 136 | 136 |
| 200 | 4 | 73 | 41 | 114 | 114 |
| 210 | 7 | 77 | 42 | 119 | 119 |
| 220 | 5 | 69 | 40 | 109 | 109 |
| 230 | 6 | 85 | 50 | 135 | 135 |
| 240 | 4 | 67 | 48 | 115 | 115 |
| 250 | 4 | 99 | 53 | 152 | 152 |
| 260 | 7 | 133 | 91 | 224 | 224 |
| 270 | 9 | 136 | 95 | 231 | 231 |
| 280 | 6 | 148 | 86 | 234 | 234 |
| 290 | 8 | 190 | 121 | 311 | 311 |
| 300 | 11 | 228 | 150 | 378 | 378 |
| 310 | 7 | 244 | 151 | 395 | 395 |
| 320 | 10 | 220 | 141 | 361 | 361 |
| 330 | 8 | 228 | 146 | 374 | 374 |
| 340 | 9 | 228 | 160 | 388 | 388 |
| 350 | 6 | 201 | 143 | 344 | 344 |
| 360 | 7 | 140 | 129 | 269 | 269 |

* Above 360 K, $R_1$ of the orthorhombic space group is larger than 10% and is not presented in the table.

**Table S4-B. Sample Peaks Violating Conditions of the *Pnma* Space Group (250 K)\***

| $(h\ k\ l)$ | $F_0^2$ | $\sigma(\ F_0^2)$ | |
|---|---|---|---|
| (0 -7 2) | 164.08 | 14.6 | observed (0 k l) reflections, |
| (0 7 -2) | 163.48 | 14. | k +1 not even |
| (0 -7 -2) | 109.39 | 10.5 | |
| (0 7 2) | 104.99 | 10.1 | |
| (0 -11 -2) | 73.29 | 7.8 | |
| (0 11 -2) | 43.9 | 4 | |
| (0 -11 -2) | 73.29 | 7.8 | |
| (0 9 -2) | 5.9 | 0.9 | |
| (0 -9 -2) | 5.8 | 1.3 | |
| ( 1 -7 0) | 379.96 | 34.2 | observed (h k 0) reflections, |
| (-1 7 0) | 373.66 | 34.4 | h not even |
| (-1 -7 0) | 414.86 | 34.8 | |
| ( 1 7 0) | 401.56 | 34.7 | |
| ( 1 -5 0) | 266.47 | 22 | |
| (-1 5 0) | 251.47 | 21.9 | |
| (-1 -5 0) | 202.78 | 17.4 | |
| ( 1 5 0) | 180.58 | 17.4 | |
| ( 3 -9 0) | 128.09 | 11.4 | |
| (-3 9 0) | 109.69 | 11.5 | |
| ( 1 11 0) | 81.69 | 7.1 | |
| (-1 -11 0) | 56.89 | 6.8 | |
| ( 1 -9 0) | 76.89 | 6.7 | |
| (-1 9 0) | 68.19 | 6.6 | |
| ( 1 9 0) | 72.49 | 6.6 | |
| (-1 -9 0) | 63.49 | 6.4 | |
| (-3 5 0) | 22.5 | 2.3 | |
| ( 3 -5 0) | 22.3 | 2.3 | |

\* The largest peak observed with respect to the the *Pnma* space group is the (4 0 0) reflections (scaled $|F_0|^2$ =10,000 and $\sigma$ ($|F_0|^2$) = 830). The intensities of the extinction violating peaks in *Pnma* are $10^2$ times weaker than the main peak.

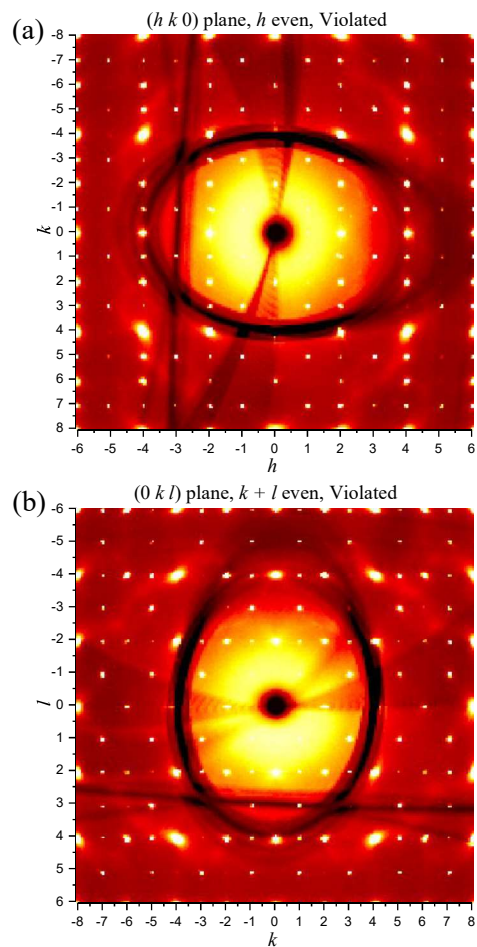

**Fig. S11.** Single-crystal X-ray diffraction reciprocal space precession images of (a) ($h$ $k$ 0) plane and (b) (0 $k$ $l$) planes of CsPbBr$_3$ at 250 K. The grid corresponds to the previously reported *Pnma* space group with unit cell dimension: $\sim\sqrt{2}\,a_P \times \sqrt{2}\,a_P \times 2a_P$ ($a$ =8.2646 Å, $b$ =11.7416 Å, $c$ =8.1707 Å). Two of the *Pnma* space group reflection conditions are 0 $k$ $l$: $k+l$ =2n and $h$ $k$ 0: $h$ =2n. Both conditions are violated by the measured data (Also see Table S2-B).

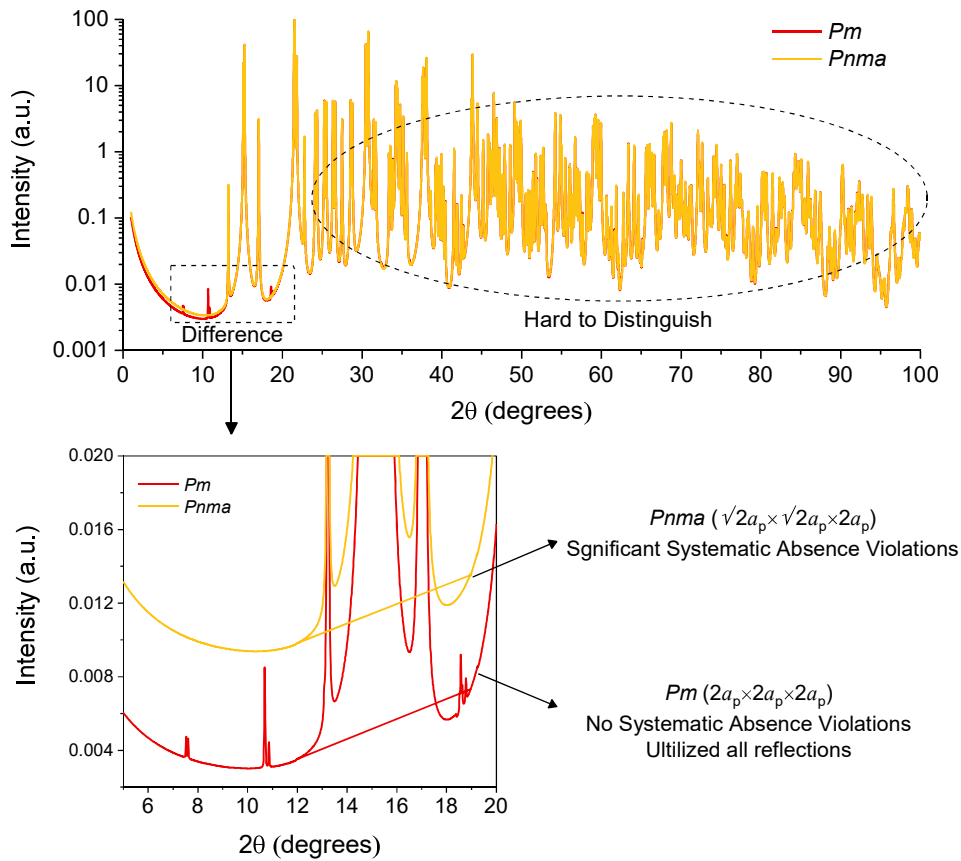

**Fig. S12-A.** The simulated powder diffraction patterns of the low-temperature models: previously reported *Pnma* structure and the newly assigned *Pm* structure. The wavelength for the simulated powder diffraction patterns is 1.54059 Å (Cu-K$\alpha$). The y-axis is the intensity on the log scale. The unit cell dimension of orthorhombic models is $\sqrt{2}\,a_P \times \sqrt{2}\,a_P \times a_P$, while the monoclinic *Pm* model is $\sim 2a_P \times 2a_P \times 2a_P$. It is very difficult to distinguish the difference between the two models at a high 2$\theta$ angle since multiple peaks merge. Note that the additional features that appear in the *Pm* model are of low intensity ($10^4$ times weaker than main peaks). To refine the structures, all reflections are utilized in the *Pm* model and reveals the real structure is polar. The orthorhombic *Pnma* model has non-indexed half-integer reflections which are the additional features seen in the *Pm* powder diffraction pattern.

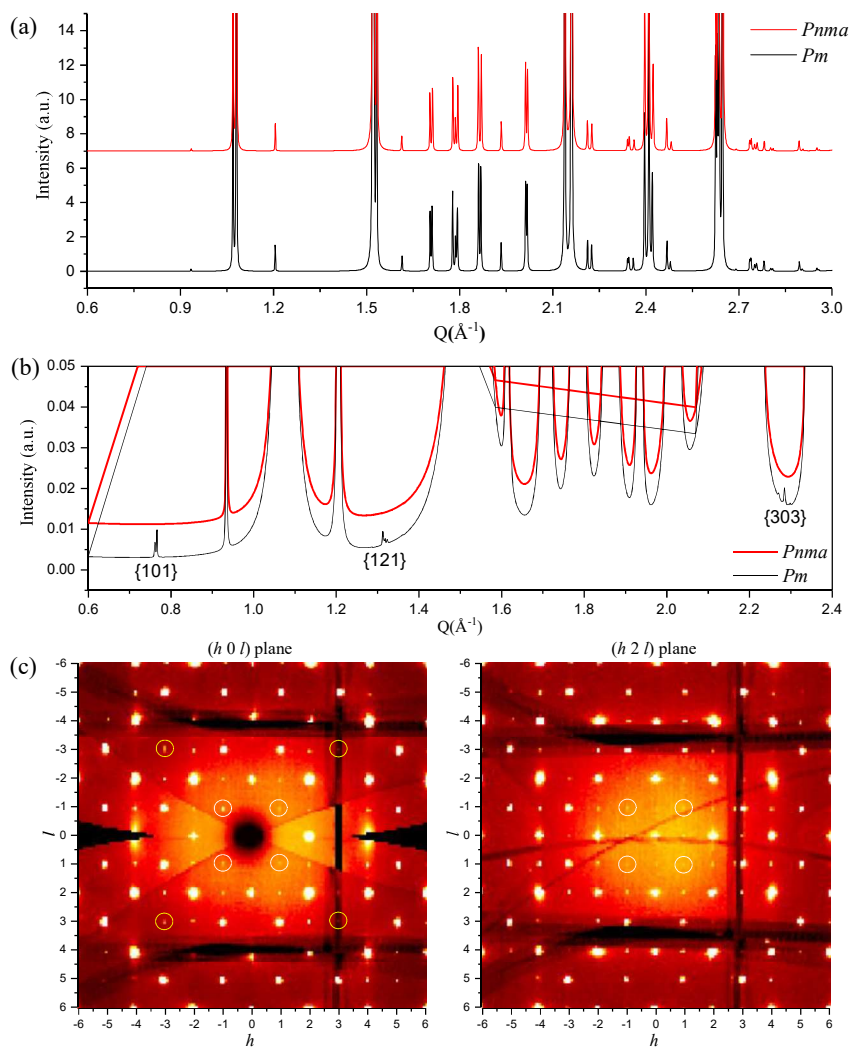

**Fig. S12-B.** (a) The simulated powder diffraction patterns of the low-temperature models: previously reported *Pnma* structure and the newly assigned *Pm* structure. The wavelength for the simulated powder diffraction patterns is 1.54059 Å (Cu-Kα). The y-axis is the intensity on the linear scale. All dominant peaks match up. (b) Additional reflections in the calculated *Pm* powder pattern compared with the pattern of the *Pnma* space group. The additional reflections in the *Pm* space group are labeled. (c) The corresponding (*h k l*) peaks of the additional features in the *Pm* space group are observed in single-crystal diffraction data (at 280 K). Note that the *b* axis is the long axis in the *Pm* structure. The circled reflection peaks are the additional peaks observed in the *Pm* structure which are labeled in Panel (b).

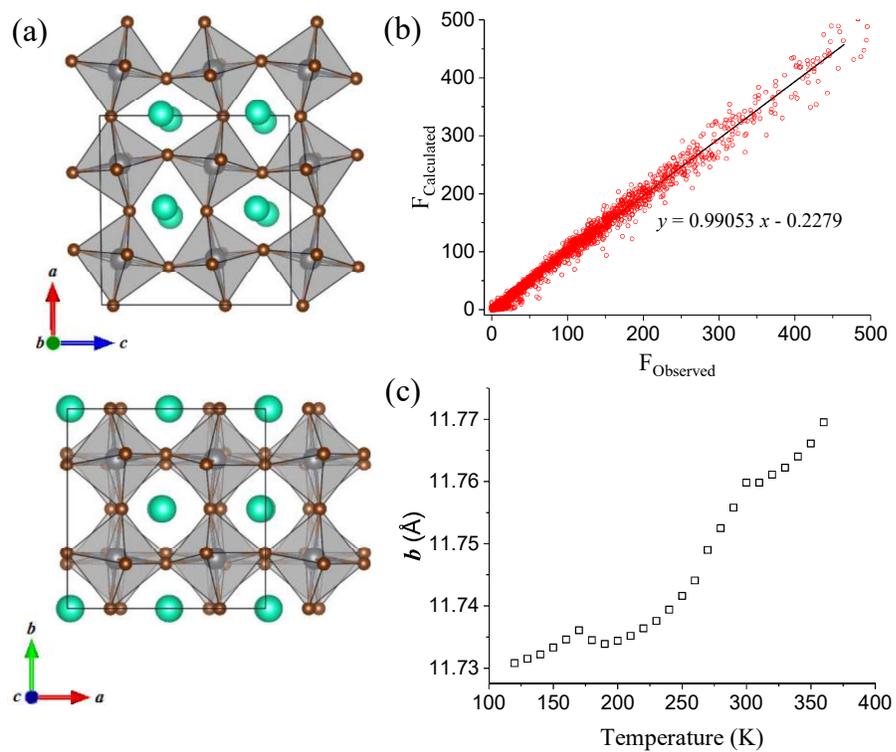

**Fig. S13-A.** (a) Solved structures from single-crystal X-ray diffraction measurements in the *Pm* space group. (b) The F$_{measured}$ vs F$_{calculated}$ is fitted by a linear function (at 120 K). (c) The lattice parameters *b* as a function of temperature.

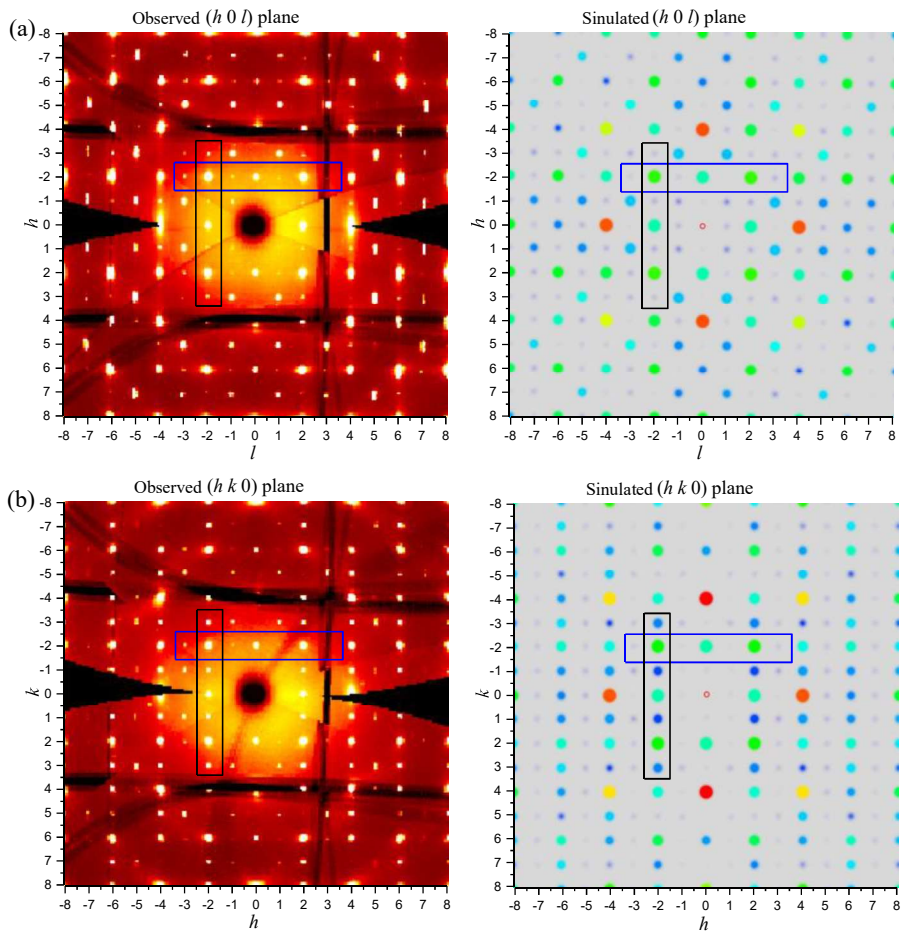

**Fig. S13-B.** Experimental reciprocal space precession images (120 K) for the (a) ($h$ 0 $l$) plane and (b) ($h$ $k$ 0) plane indexed with a $2a_P \times 2a_P \times 2a_P$ cell compared to the simulated pattern using the *Pm* crystal structure solution ($2a_P \times 2a_P \times 2a_P$). The *b* axis is the long axis. Observe the qualitative similarity in the weak peaks representing the doubled cell relative to the simple cubic cell ($a_P \times a_P \times a_P$). Rectangles with the same color cover equivalent regions in measured and simulated images. In the calculation, the high-intensity peaks are in the red region of the color spectrum and the low-intensity peaks are on the blue end of the spectrum. The size of the peaks shown also indicates their intensities.

**Table S5.** Structural Parameters from CsPbBr$_3$ at 450 K in *Im*-3 Space Group

| Atoms | x | y | z | U$_{eq}$ (Å$^3 \times 10^3$) | | |
|---|---|---|---|---|---|---|
| Pb1 | 7500 | 2500 | 2500 | 37.0(3) | | |
| Cs1 | 5000 | 5000 | 5000 | 117.1(12) | | |
| Cs2 | 5000 | 0 | 5000 | 118.5(14) | | |
| Br1 | 5000 | 2432.8(16) | 2565.2(16) | 158.4(13) | | |
| *U$_{ij}$* (Pb1) | 37.0(3) | 37.0(3) | 37.0(3) | -0.02(5) | -0.02(5) | -0.02(5) |
| *U$_{ij}$* (Cs1) | 118.8(15) | 102.9(13) | 129.5(18) | 0 | 0 | 0 |
| *U$_{ij}$* (Cs2) | 118.5(14) | 118.5(14) | 118.5(14) | 0 | 0 | 0 |
| *U$_{ij}$* (Br1) | 225(4) | 219(4) | 31.3(6) | 0 | 0 | 11.1(12) |

Space Group: *Im*-3
*a* = 11.7444 (3) Å, Dx = 4.755 g/cm$^3$
Measurement Temperature: 450 K
Crystal Dimensions (diameter): ~50 μm
Wavelength: 0.41328 Å
2θ range for data collection: 2.852° to 44.138°
Index ranges: -18 ≤ h ≤ 18, -18 ≤ k ≤ 18, -18 ≤ l ≤ 13
Reflections collected: 35245
EXTI extinction parameter: 0.0107(11)
Independent reflections: 863
Number of fitting parameters: 14
Largest diff. peak/hole: 1.32(Cs2)/-1.20(Br1) e Å$^{-3}$
R$_1$ = 3.30 %, wR$_2$ = 8.40 %, Goodness of Fit = 1.012

[*]Atomic displacement parameters $U_{ij}$ (Å$^2 \times 10^3$) are in the order $U_{11}$, $U_{22}$, $U_{33}$, $U_{23}$, $U_{13}$, $U_{12}$

**Table S6.** Structural Parameters from CsPbBr$_3$ at 380 K in $P2_1/m$ Space Group

| Atoms | x | y | z | U$_{eq}$ (Å$^3$×10$^3$) | | |
|---|---|---|---|---|---|---|
| Pb1 | 5000 | 0 | 0 | 28.35(19) | | |
| Pb2 | 0 | 0 | 0 | 28.42(19) | | |
| Pb3 | 0 | 0 | 5000 | 28.29(19) | | |
| Pb3 | 5000 | 0 | 5000 | 28.44(19) | | |
| Cs1 | 7611.1(19) | 2500 | 2410.4(11) | 92.1(6) | | |
| Cs2 | 7582.8(12) | 2500 | 7374.8(18) | 87.1(5) | | |
| Cs3 | 2625.9(2) | 2500 | 7435.7(11) | 91.0(6) | | |
| Cs4 | 2572.4(13) | 2500 | 2386.1(10) | 96.2(6) | | |
| Br1 | 4567(16) | -182.9(14) | 2497.2(10) | 80.7(4) | | |
| Br2 | -433.5(16) | -185.5(14) | 7496.7(10) | 80.1(4) | | |
| Br3 | 7494.6(10) | 188(14) | 4562.6(10) | 79.3(4) | | |
| Br4 | 4794.2(3) | 2500 | 4830.3(3) | 92.9(8) | | |
| Br5 | 7504.0(11) | -175.3(14) | 439.1(15) | 80.3(4) | | |
| Br6 | -201.7(3) | 2500 | -168.4(3) | 97.1(9) | | |
| Br7 | 5157.5(3) | 2500 | 197.6(3) | 97.2(9) | | |
| Br8 | 182.7(3) | 2500 | 5203.9(3) | 91.1(8) | | |
| $U_{ij}$ (Pb1) | 30.6(3) | 26.4(3) | 28.1(3) | 0.83(8) | 1.5(3) | -0.15(8) |
| $U_{ij}$ (Pb2) | 31.1(3) | 26.9(3) | 27.3(3) | 0.06(8) | 1.6(3) | -0.73(9) |
| $U_{ij}$ (Pb3) | 30.7(3) | 26.6(3) | 27.6(3) | 0.77(8) | 1.6(3) | -0.10(8) |
| $U_{ij}$ (Pb4) | 30.9(3) | 26.6(3) | 27.8(3) | 0.15(8) | 1.6(3) | -0.76(8) |
| $U_{ij}$ (Cs1) | 149.7(18) | 61.3(7) | 65.2(7) | 0 | -1.1(10) | 0 |
| $U_{ij}$ (Cs2) | 68.8(8) | 66.0(8) | 126.6(13) | 0 | -8.1(9) | 0 |
| $U_{ij}$ (Cs3) | 146.9(17) | 64.6(8) | 61.6(6) | 0 | -6.7(9) | 0 |
| $U_{ij}$ (Cs4) | 69.1(9) | 62.7(8) | 156.7(17) | 0 | -17.2(10) | 0 |
| $U_{ij}$ (Br1) | 97.4(10) | 118.5(10) | 26.1(4) | 0.7(5) | 2.5(5) | -11.8(8) |
| $U_{ij}$ (Br2) | 97.3(10) | 117.5(10) | 25.5(4) | -3.1(5) | 2.4(5) | -10.9(8) |
| $U_{ij}$ (Br3) | 27.8(4) | 115.8(10) | 94.3(9) | 15.1(7) | 2.4(5) | 2.2(5) |
| $U_{ij}$ (Br4) | 134(2) | 27.2(6) | 117.1(18) | 0 | -0.8(15) | 0 |
| $U_{ij}$ (Br5) | 27.6(5) | 118.9(10) | 94.4(10) | 8.6(7) | 2.1(5) | -0.1(5) |
| $U_{ij}$ (Br6) | 149(2) | 27.5(6) | 114.8(18) | 0 | 1.3(16) | 0 |
| $U_{ij}$ (Br7) | 125(2) | 24.7(6) | 142(2) | 0 | 3.2(16) | 0 |
| $U_{ij}$ (Br8) | 112.6(19) | 26.6(6) | 134.1(19) | 0 | 5.6(15) | 0 |

Space Group: $P2_1/m$
$a$ = 11.6630(4) Å, $b$ = 11.7796(5) Å, $c$ = 11.6664(5) Å, $\beta$ = 90.0570(10)°, Dx = 4.806 g/cm$^3$
Measurement Temperature: 380 K
Crystal Dimensions (diameter): ~50 μm
Wavelength: 0.41328 Å
2θ range for data collection: 2.01° to 44.144°
Index ranges: -18 ≤ h ≤ 13, -18 ≤ k ≤ 18, -18 ≤ l ≤ 18
Reflections collected: 68421
Twin law: 1 0 0 0 -1 0 0 0 -1 2
BASF parameter: 0.265(4)
EXTI extinction parameter: 0.0118(12)
Independent reflections: 7966
Number of fitting parameters: 111
Largest diff. peak/hole: 4.89(Pb4)/-6.8(Cs4) e Å$^{-3}$
R$_1$ = 7.22 %, wR$_2$ = 32.47 %, Goodness of Fit = 1.006

*Atomic displacement parameters $U_{ij}$ (Å$^2$×10$^3$) are in the order $U_{11}$, $U_{22}$, $U_{33}$, $U_{23}$, $U_{13}$, $U_{12}$.
** The Pseudomerohedry twin fraction components are 0.735(5) and 0.265(5).

**Table S7.** Structural Parameters from CsPbBr$_3$ at 340 K in $P2_1/m$ Space Group

| Atoms | x | y | z | U$_{eq}$ (Å$^3$×10$^3$) | | |
|---|---|---|---|---|---|---|
| Pb1 | 5000 | 0 | 0 | 28.14(13) | | |
| Pb2 | 0 | 0 | 0 | 28.13(13) | | |
| Pb3 | 0 | 0 | 5000 | 28.09(13) | | |
| Pb3 | 5000 | 0 | 5000 | 28.16(13) | | |
| Cs1 | 7656.1(18) | 2500 | 2383.2(10) | 84.9(5) | | |
| Cs2 | 7612.6(12) | 2500 | 7325.3(16) | 79.7(4) | | |
| Cs3 | 2674.1(18) | 2500 | 7403.9(10) | 83.5(4) | | |
| Cs4 | 2604.8(13) | 2500 | 2332.6(18) | 88.8(4) | | |
| Br1 | 4559.7(13) | -232.7(12) | 2497.5(8) | 70.9(4) | | |
| Br2 | -441.1(13) | -235.7(12) | 7498.3(8) | 70.6(4) | | |
| Br3 | 7497.3(10) | 236.8(12) | 4562.3(12) | 69.9(3) | | |
| Br4 | 4749.8(3) | 2500 | 4789.6(2) | 81.4(6) | | |
| Br5 | 7501.4(11) | -227.3(12) | 441.7(12) | 70.5(3) | | |
| Br6 | -252.3(3) | 2500 | -215.9(2) | 85.0(6) | | |
| Br7 | 5195.8(2) | 2500 | 238.1(3) | 86.4(6) | | |
| Br8 | 223.2(2) | 2500 | 5246.9(2) | 79.6(6) | | |
| $U_{ij}$ (Pb1) | 30.8(2) | 25.4(2) | 28.2(2) | 1.10(10) | 1.04(16) | -0.21(8) |
| $U_{ij}$ (Pb2) | 31.0(2) | 25.7(2) | 27.6(2) | 0.17(10) | 1.16(16) | -1.18(8) |
| $U_{ij}$ (Pb3) | 30.9(2) | 25.7(2) | 27.7(2) | 1.05(10) | 1.20(16) | -0.13(8) |
| $U_{ij}$ (Pb4) | 31.0(2) | 25.3(2) | 28.2(2) | 0.17(10) | 1.02(16) | -0.94(8) |
| $U_{ij}$ (Cs1) | 139.5(14) | 56.9(6) | 58.3(5) | 0 | -3.3(7) | 0 |
| $U_{ij}$ (Cs2) | 63.9(8) | 60.8(6) | 114.2(10) | 0 | -4.2(7) | 0 |
| $U_{ij}$ (Cs3) | 130.4(13) | 60.6(6) | 59.5(6) | 0 | -6.6(7) | 0 |
| $U_{ij}$ (Cs4) | 64.5(9) | 59.7(6) | 142.0(13) | 0 | -14.3(8) | 0 |
| $U_{ij}$ (Br1) | 89.5(9) | 97.5(8) | 25.7(3) | 1.7(4) | 1.5(4) | -11.6(7) |
| $U_{ij}$ (Br2) | 90.9(9) | 96.2(8) | 24.5(3) | -1.3(4) | 1.4(4) | -11.5(7) |
| $U_{ij}$ (Br3) | 28.9(5) | 95.6(8) | 85.2(7) | 14.2(6) | 2.0(4) | 1.9(5) |
| $U_{ij}$ (Br4) | 121.2(16) | 23.2(5) | 99.7(13) | 0 | -8.0(12) | 0 |
| $U_{ij}$ (Br5) | 29.5(5) | 98.1(8) | 83.9(7) | 9.3(6) | 1.3(4) | -1.4(5) |
| $U_{ij}$ (Br6) | 134.1(18) | 24.4(5) | 96.6(12) | 0 | -9.0(13) | 0 |
| $U_{ij}$ (Br7) | 111.5(16) | 21.0(5) | 126.5(16) | 0 | -0.9(13) | 0 |
| $U_{ij}$ (Br8) | 97.3(14) | 23.2(5) | 118.2(15) | 0 | -1.0(12) | 0 |

Space Group: $P2_1/m$
$a$ = 11.6457(4) Å, $b$ = 11.7640(4) Å, $c$ = 11.6497(4) Å, $\beta$ = 90.1520(10)°, Dx = 4.826 g/cm$^3$
Measurement Temperature: 340 K
Crystal Dimensions (diameter): ~50 μm
Wavelength: 0.41328 Å
2θ range for data collection: 2.012° to 44.134°
Index ranges: -13 ≤ h ≤ 18, -18 ≤ k ≤ 18, -18 ≤ l ≤ 18
Reflections collected: 70053
Twin law: 1 0 0 0 -1 0 0 0 -1 2
BASF parameter: 0.192(3)
EXTI extinction parameter: 0.0154(9)
Independent reflections: 7584
Number of fitting parameters: 111
Largest diff. peak/hole: 3.36(Cs2)/-5.06(Cs1) e Å$^{-3}$
R$_1$ = 6.10 %, wR$_2$ = 22.64 %, Goodness of Fit = 1.003

*Atomic displacement parameters $U_{ij}$ (Å$^2$×10$^3$) are in the order $U_{11}$, $U_{22}$, $U_{33}$, $U_{23}$, $U_{13}$, $U_{12}$.
** The Pseudomerohedry twin fraction components are 0.808(3) and 0.192(3).

**Table S8.** Structural Parameters from CsPbBr₃ at 280 K in *Pm* Space Group

| Atoms | x | y | z | U$_{eq}$ | U$_{11}$ | U$_{22}$ | U$_{33}$ | U$_{23}$ | U$_{13}$ | U$_{12}$ |
|---|---|---|---|---|---|---|---|---|---|---|
| Pb1 | 2531.5(9) | 2498.7(5) | 7531.1(8) | 26.44(18) | 30.9(4) | 22.6(3) | 25.8(3) | -1.25(15) | 1.2(2) | 0.05(14) |
| Pb2 | 7531.3(9) | 2499.1(5) | 7527.8(8) | 26.48(18) | 30.8(4) | 23.0(3) | 25.6(3) | -0.22(15) | 1.1(2) | 1.03(14) |
| Pb3 | 7532.1(9) | 2499.0(5) | 2527.6(8) | 26.47(18) | 30.9(4) | 23.0(3) | 25.5(3) | -1.13(15) | 1.1(2) | 0.31(14) |
| Pb4 | 2531.2(9) | 2498.8(5) | 2531.1(8) | 26.37(18) | 30.8(4) | 22.6(3) | 25.7(3) | -0.14(15) | 1.3(2) | 1.25(14) |
| Cs1 | 10255(5) | 0 | 4909(4) | 71.5(14) | 115(4) | 50.1(11) | 49.5(16) | 0 | -6.8(19) | 0 |
| Cs2 | 5251(5) | 0 | 9905(4) | 72.0(14) | 118(4) | 48.8(11) | 48.9(15) | 0 | -5.9(19) | 0 |
| Cs3 | 10145(5) | 0 | 9890(6) | 85.7(18) | 50(3) | 50.8(13) | 156(5) | 0 | -13(3) | 0 |
| Cs4 | 5170(5) | 0 | 4846(6) | 76.7(14) | 50(3) | 52.9(12) | 127(4) | 0 | -4(2) | 0 |
| Cs5 | 4893(5) | 5000 | 5285(2) | 74.2(12) | 74(3) | 55.4(13) | 94(3) | 0 | -8(2) | 0 |
| Cs6 | -95(5) | 5000 | 277(5) | 71.3(11) | 72(3) | 56.2(12) | 86(2) | 0 | -3.9(19) | 0 |
| Cs7 | 4821(6) | 5000 | 167(5) | 81.1(16) | 117(5) | 58.4(14) | 68(2) | 0 | -3(2) | 0 |
| Cs8 | -154(6) | 5000 | 5189(5) | 84.0(18) | 136(6) | 56.4(14) | 59(2) | 0 | -5(2) | 0 |
| Br1 | 2967(5) | 2239(3) | 5013(4) | 62.6(9) | 79(3) | 83.6(16) | 25.2(8) | -1.8(11) | -1.4(12) | 10.7(15) |
| Br2 | 2289(8) | 5000 | 7239(8) | 80(2) | 87(5) | 24.4(13) | 129(6) | 0 | -14(4) | 0 |
| Br3 | 7280(7) | 5000 | 2229(7) | 73.8(19) | 81(5) | 26.5(12) | 114(5) | 0 | -8(3) | 0 |
| Br4 | 7977(5) | 2247(2) | 9(4) | 60.9(9) | 75(3) | 82.8(16) | 24.5(8) | -0.5(11) | -0.8(11) | 8.7(15) |
| Br5 | 30(5) | 2771(2) | 2959(4) | 61.7(8) | 28(2) | 79.8(16) | 78(2) | -12.8(13) | 0.2(13) | -2.5(12) |
| Br6 | 7259(9) | 0 | 7254(7) | 84(2) | 128(7) | 27.9(14) | 96(5) | 0 | -16(4) | 0 |
| Br7 | 5023(5) | 2242(2) | 2065(4) | 61.7(8) | 31(2) | 80.8(16) | 73(2) | -8.5(13) | 2.8(14) | 0.6(13) |
| Br8 | 7072(5) | 2769(2) | 5003(4) | 63.8(10) | 90(3) | 80.1(15) | 21.0(8) | 1.5(10) | 3.9(12) | 13.2(15) |
| Br9 | 2062(5) | 2758(2) | 5(4) | 63.2(10) | 85(3) | 82.3(15) | 21.9(8) | 0.5(10) | 4.2(12) | 8.7(15) |
| Br10 | 5027(5) | 2754(2) | 7969(4) | 61.2(8) | 29(2) | 79.9(16) | 75(2) | -7.8(13) | -0.7(13) | -0.1(12) |
| Br11 | 2261(9) | 0 | 2267(8) | 82(2) | 115(7) | 28.2(14) | 104(5) | 0 | -11(4) | 0 |
| Br12 | 24(5) | 2237(3) | 7069(4) | 62.6(8) | 32(2) | 82.8(16) | 73(2) | -10.9(13) | 3.2(14) | 1.8(13) |
| Br13 | 2803(7) | 5000 | 2735(6) | 64.0(17) | 104(5) | 15.9(9) | 72(3) | 0 | 1(3) | 0 |
| Br14 | 7807(7) | 5000 | 7739(6) | 65.8(17) | 113(6) | 16.1(9) | 68(3) | 0 | -2(3) | 0 |
| Br15 | 2736(8) | 0 | 7784(7) | 75.4(19) | 105(6) | 15.9(10) | 105(5) | 0 | -1(4) | 0 |
| Br16 | 7755(7) | 0 | 2782(7) | 70.3(18) | 94(5) | 19.8(10) | 97(4) | 0 | 5(3) | 0 |

Space Group: *Pm*
a = 11.6324(5) Å, b = 11.7525(6) Å, c = 11.6368(6) Å, β = 89.663(10)°, Dx = 4.842 g/cm³
Measurement Temperature: 280 K
Crystal Dimensions (diameter): ~50 μm
Wavelength: 0.41328 Å
2θ range for data collection: 2.014° to 37.07°
Index ranges: -13 ≤ h ≤ 16, -17 ≤ k ≤ 17, -17 ≤ l ≤ 17
Reflections collected: 61776
Twin law: 1 0 0 0 -1 0 0 0 -1 2
BASF parameter: 0.136(3)
EXTI extinction parameter: 0.0118(8)
Flack parameter: 0.55(8)
Independent reflections: 10493
Number of fitting parameters: 207
Largest diff. peak/hole: 2.69(Br16)/-4.97(Cs2) e Å⁻³
R$_1$ = 5.13 %, wR$_2$ = 20.92 %, Goodness of Fit = 1.113

*Unit of Atomic displacement parameters is Å²×10³.
**The Pseudomerohedry twin fraction components are 0.864(7) and 0.136(3). The Racemic twin fraction components are 0.45(8) and 0.55(8).

**Table S9.** Structural Parameters from CsPbBr$_3$ at 250 K in *Pm* Space Group

| Atoms | x | y | z | U$_{eq}$ | U$_{11}$ | U$_{22}$ | U$_{33}$ | U$_{23}$ | U$_{13}$ | U$_{12}$ |
|---|---|---|---|---|---|---|---|---|---|---|
| Pb1 | 2539.9(15) | 2501.6(4) | 7542.8(10) | 23.6(2) | 28.1(4) | 20.3(3) | 22.5(3) | 0.01(13) | 1.2(2) | 1.31(14) |
| Pb2 | 7540.9(14) | 2501.3(4) | 7542.4(10) | 23.6(2) | 28.0(4) | 20.3(3) | 22.5(3) | -1.15(13) | 1.0(2) | 0.03(14) |
| Pb3 | 7540.3(14) | 2501.9(4) | 2542.7(10) | 23.6(2) | 28.0(4) | 20.3(3) | 22.6(3) | -0.23(13) | 1.0(2) | 1.17(14) |
| Pb4 | 2541.3(14) | 2501.5(4) | 2542.8(10) | 23.6(2) | 28.0(4) | 20.3(3) | 22.5(3) | -1.38(13) | 1.1(2) | 0.03(14) |
| Cs1 | 9828.2(7) | 0 | 4879(4) | 75.5(11) | 119(4) | 51.1(12) | 57.0(18) | 0 | 0.5(17) | 0 |
| Cs2 | 4815.4(7) | 0 | 9854.1(4) | 75.3(12) | 123(4) | 53.5(12) | 49.1(15) | 0 | 0.3(16) | 0 |
| Cs3 | 9888.2(6) | 0 | 9736.4(5) | 63.6(7) | 70(2) | 50.8(11) | 70.2(17) | 0 | -6.1(13) | 0 |
| Cs4 | 4894.9(6) | 0 | 4742.3(5) | 61.4(7) | 69(2) | 50.3(11) | 64.7(16) | 0 | -4.1(12) | 0 |
| Cs5 | 5174.8(6) | 5000 | 5189.9(5) | 79.1(13) | 41.3(14) | 46.1(11) | 150(4) | 0 | -6.0(18) | 0 |
| Cs6 | 179.3(6) | 5000 | 239(5) | 71.5(10) | 43.7(13) | 47.2(11) | 124(3) | 0 | -1.2(15) | 0 |
| Cs7 | 5290.9(7) | 5000 | 163.9(5) | 63.9(8) | 94(2) | 44.3(10) | 53.7(15) | 0 | -6.7(13) | 0 |
| Cs8 | 287.2(7) | 5000 | 5161.9(5) | 63.7(8) | 94(2) | 45.2(10) | 51.9(14) | 0 | -4.9(13) | 0 |
| Br1 | 2987.8(6) | 2765.3(2) | 5052.8(4) | 55.5(6) | 69.1(16) | 80.3(14) | 17.2(7) | -2.8(8) | 0.3(7) | 10.9(11) |
| Br2 | 2256.7(7) | 5000 | 7877.3(7) | 74.5(15) | 142(5) | 29.4(13) | 51.2(18) | 0 | -15(2) | 0 |
| Br3 | 7274.1(7) | 5000 | 2877.2(6) | 74.0(14) | 140(5) | 29.3(13) | 52.8(19) | 0 | -13(2) | 0 |
| Br4 | 7989.2(6) | 2765.9(2) | 50.4(4) | 55.6(6) | 67.9(16) | 79.2(14) | 18.0(7) | -3.3(8) | 0.1(7) | 11.2(11) |
| Br5 | 24.4(6) | 2773.5(2) | 3020.8(4) | 53.8(6) | 27.4(10) | 73.7(14) | 60.3(14) | -4.2(10) | 4.3(8) | 1.6(9) |
| Br6 | 7308.8(8) | 0 | 7871.7(6) | 71.5(13) | 122(4) | 25.5(12) | 67(2) | 0 | -7(2) | 0 |
| Br7 | 5035.5(6) | 2218.1(2) | 2104.2(4) | 55.9(6) | 24.2(10) | 69.6(13) | 74.0(17) | -13.1(11) | -1.2(8) | -0.5(9) |
| Br8 | 7062.5(6) | 2215.9(2) | 5059.5(4) | 55.1(6) | 73.4(17) | 67.8(13) | 24.3(8) | 0.9(8) | 2.7(8) | 8.8(11) |
| Br9 | 2064.9(6) | 2218.6(2) | 62.1(4) | 54.9(6) | 73.5(17) | 68.3(13) | 22.9(8) | 1.5(8) | 2.8(8) | 7.9(11) |
| Br10 | 5024.6(6) | 2780.2(2) | 8018.4(4) | 54.8(6) | 27.3(10) | 75.8(14) | 61.4(14) | -9.1(10) | 4.3(8) | 1.9(9) |
| Br11 | 2310.5(8) | 0 | 2873.9(6) | 68.7(12) | 115(4) | 25.6(12) | 65(2) | 0 | -2(2) | 0 |
| Br12 | 33.4(6) | 2234.3(2) | 7094.1(4) | 54.7(6) | 24.2(10) | 68.4(12) | 71.4(16) | -9.2(11) | -1.1(8) | -0.5(9) |
| Br13 | 2791.5(8) | 5000 | 2310.5(5) | 67.7(12) | 59(2) | 14.4(9) | 130(4) | 0 | 0(2) | 0 |
| Br14 | 7787.7(8) | 5000 | 7302.5(5) | 71.1(13) | 60(2) | 13.4(9) | 140(4) | 0 | -3(2) | 0 |
| Br15 | 2825.1(8) | 0 | 7364.1(5) | 59.2(10) | 91(3) | 11.8(8) | 74(2) | 0 | -0.4(18) | 0 |
| Br16 | 7833.5(8) | 0 | 2374.8(5) | 58.4(10) | 87(3) | 11.0(8) | 77(2) | 0 | -0.9(18) | 0 |

Space Group: *Pm*
$a$ = 11.6205(4) Å, $b$ = 11.7416(5) Å, $c$ = 11.6238(5) Å, $\beta$ = 89.357(10)°, Dx = 4.857 g/cm$^3$
Measurement Temperature: 250 K
Crystal Dimensions (diameter): ~50 µm
Wavelength: 0.41328 Å
2θ range for data collection: 2.018° to 37.072°
Index ranges: -17 ≤ h ≤ 13, -17 ≤ k ≤ 17, -17 ≤ l ≤ 17
Reflections collected: 63103
Twin law: 1 0 0 0 -1 0 0 0 -1 2
BASF parameter: 0.136(3)
Flack parameter: 0.57(7)
EXTI extinction parameter: 0.0093 (9)
Independent reflections: 11167
Number of fitting parameters: 207
Largest diff. peak/hole: 7.52(Pb3)/-6.37(Cs7) e Å$^{-3}$
R$_1$ = 6.86 %, wR$_2$ = 25.56 %, Goodness of Fit = 1.120

*Unit of Atomic displacement parameters is Å$^2$×10$^3$.
**The Pseudomerohedry twin fraction components are 0.9309(19) and 0.0691(19). The Racemic twin fraction components are 0.43(8) and 0.57(7).

**Table S10.** Structural Parameters from CsPbBr$_3$ at 120 K in *Pm* Space Group

| Atoms | x | y | z | U$_{eq}$ | U$_{11}$ | U$_{22}$ | U$_{33}$ | U$_{23}$ | U$_{13}$ | U$_{12}$ |
|---|---|---|---|---|---|---|---|---|---|---|
| Pb1 | 2530.2(17) | 2499.7(4) | 7530.7(11) | 22.0(2) | 26.7(4) | 18.8(4) | 20.1(4) | 0.10(15) | 1.7(2) | 1.21(16) |
| Pb2 | 7529.3(16) | 2497.9(4) | 7529.9(11) | 21.9(2) | 26.6(4) | 18.6(4) | 20.4(4) | -1.22(15) | 1.7(2) | -0.15(17) |
| Pb3 | 7530.9(17) | 2499.7(4) | 2529.3(11) | 22.0(2) | 26.6(4) | 18.6(4) | 20.3(4) | -0.17(15) | 1.4(2) | 1.10(16) |
| Pb4 | 2529.7(16) | 2497.8(4) | 2530.8(11) | 22.0(2) | 26.6(4) | 18.6(4) | 20.4(4) | -1.35(15) | 1.6(2) | -0.21(17) |
| Cs1 | 10319(7) | 0 | 4867(5) | 60.3(8) | 73(2) | 44.4(11) | 55.0(15) | 0 | -1.9(12) | 0 |
| Cs2 | 5306(7) | 0 | 9867(5) | 59.7(8) | 74(2) | 43.8(11) | 53.4(15) | 0 | 0.0(12) | 0 |
| Cs3 | 10187(7) | 0 | 9811(6) | 71.0(11) | 39.7(15) | 47.1(13) | 134(4) | 0 | -8.1(18) | 0 |
| Cs4 | 5194(7) | 0 | 4771(6) | 63.6(9) | 42.1(15) | 47.8(12) | 105(3) | 0 | -1.6(15) | 0 |
| Cs5 | 4869(7) | 5000 | 5345(5) | 56.4(7) | 61.7(19) | 42.0(11) | 64.8(18) | 0 | -0.5(12) | 0 |
| Cs6 | -123(7) | 5000 | 331(5) | 55.1(7) | 61.9(19) | 41.4(10) | 60.0(17) | 0 | -1.5(12) | 0 |
| Cs7 | 4768(7) | 5000 | 188(5) | 64.3(8) | 118(3) | 40.7(11) | 40.7(13) | 0 | -1.9(15) | 0 |
| Cs8 | -239(7) | 5000 | 5215(5) | 62.0(8) | 121(3) | 43.1(11) | 33.1(11) | 0 | -1.0(13) | 0 |
| Br1 | 3014(7) | 2218(2) | 5005(4) | 48.1(6) | 54.8(14) | 67.0(14) | 21.7(8) | 1.1(8) | 2.8(7) | 5.2(10) |
| Br2 | 2250(8) | 5000 | 7284(7) | 63.7(13) | 113(4) | 12.3(9) | 40.6(17) | 0 | 4.3(16) | 0 |
| Br3 | 7255(8) | 5000 | 2271(7) | 60.5(12) | 108(4) | 11.2(9) | 43.5(17) | 0 | 4.8(17) | 0 |
| Br4 | 8012(7) | 2215(2) | 3(4) | 48.5(6) | 55.4(15) | 66.1(14) | 22.8(8) | 0.0(8) | 3.7(8) | 7.2(10) |
| Br5 | 47(7) | 2802(2) | 2966(5) | 48.7(6) | 27.3(12) | 60.7(13) | 52.3(14) | -10.1(10) | 4.4(8) | 0.4(9) |
| Br6 | 7197(8) | 0 | 7297(7) | 62.0(12) | 70(3) | 9.3(9) | 94(3) | 0 | 1(2) | 0 |
| Br7 | 5031(7) | 2209(2) | 2033(5) | 48.0(6) | 22.0(11) | 62.0(14) | 63.3(16) | -4.2(10) | -0.7(8) | 0.0(9) |
| Br8 | 7065(7) | 2797(2) | 4989(5) | 49.3(6) | 70.9(18) | 58.4(13) | 18.3(7) | -2.1(8) | 0.4(8) | 10.3(11) |
| Br9 | 2062(7) | 2801(2) | -8(5) | 50.0(6) | 72.4(18) | 58.9(13) | 18.3(7) | -1.7(8) | 0.9(8) | 9.4(11) |
| Br10 | 5045(7) | 2783(2) | 7974(5) | 47.6(6) | 27.4(12) | 60.2(13) | 49.3(13) | -5.0(9) | 4.7(8) | 1.1(9) |
| Br11 | 2207(8) | 0 | 2291(7) | 61.2(11) | 68(3) | 10.4(9) | 90(3) | 0 | 3(2) | 0 |
| Br12 | 32(7) | 2203(2) | 7030(5) | 48.6(6) | 21.8(11) | 63.0(14) | 64.1(16) | -9.5(11) | -0.8(8) | 0.9(9) |
| Br13 | 2834(8) | 5000 | 2798(5) | 56.5(12) | 74(3) | 27.0(13) | 71(3) | 0 | -1(2) | 0 |
| Br14 | 7807(9) | 5000 | 7805(5) | 56.6(12) | 77(3) | 28.5(14) | 74(3) | 0 | -7(2) | 0 |
| Br15 | 2761(8) | 0 | 7861(6) | 55.4(11) | 85(3) | 24.2(13) | 83(3) | 0 | -16(2) | 0 |
| Br16 | 7780(8) | 0 | 2851(6) | 54.2(11) | 82(3) | 23.1(12) | 82(3) | 0 | -15(2) | 0 |

Space Group: *Pm*
$a = 11.6096(6)$ Å, $b = 11.7308(6)$ Å, $c = 11.6122(6)$ Å, $\beta = 89.062(10)°$, Dx = 4.871 g/cm$^3$
Measurement Temperature: 120 K
Crystal Dimensions (diameter): ~50 μm
Wavelength: 0.41328 Å
2θ range for data collection: 2.018° to 37.072°
Index ranges: -17 ≤ h ≤ 13, -17 ≤ k ≤ 17, -17 ≤ l ≤ 17
Reflections collected: 60526
Twin law: 1 0 0 0 -1 0 0 0 -1 2
BASF parameter: 0.0794(17)
Flack parameter: 0.52(7)
EXTI extinction parameter: 0.0109 (11)
Independent reflections: 10996
Number of fitting parameters: 207
Largest diff. peak/hole: 8.05(Pb3)/-5.49(Cs4) e Å$^{-3}$
R$_1$ = 7.78 %, wR$_2$ = 28.60 %, Goodness of Fit = 1.144

*Unit of Atomic displacement parameters is Å$^2$×10$^3$.
**The Pseudomerohedry twin fraction components are 0.9206(17) and 0.0794(17). The Racemic twin fraction components are 0.48(7) and 0.52(7).

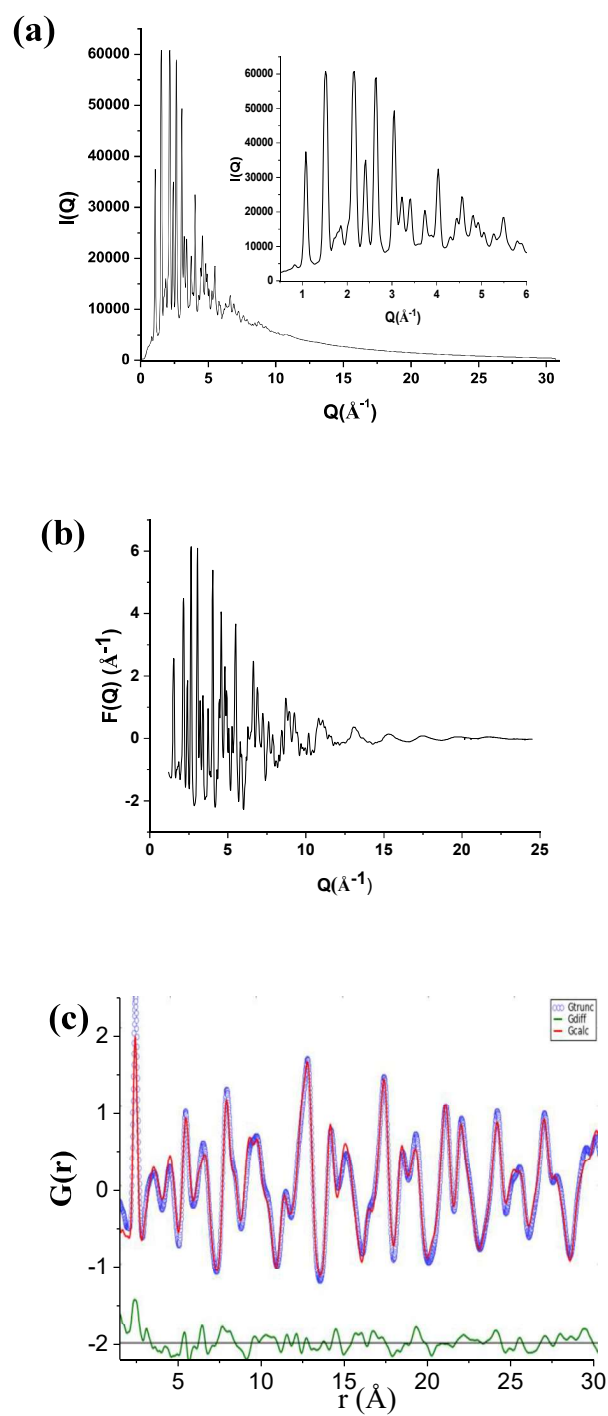

**Fig. S14.** (a) Raw X-ray scattering data (280 K) without background subtraction (expanded in inset) and (b) reduced scattering data F(Q) at 280 K. A representative fit of the PDF data in real space between 2 Å and 30 Å (c).

**Fig. S15.** (Top) Fits of to the Br K-edge x-ray absorption fine structure data at 70 K for the Br-Pb, Br-Cs, and Br-Br shells and (Bottom) at 125 K for a single Br-Pb shell only.

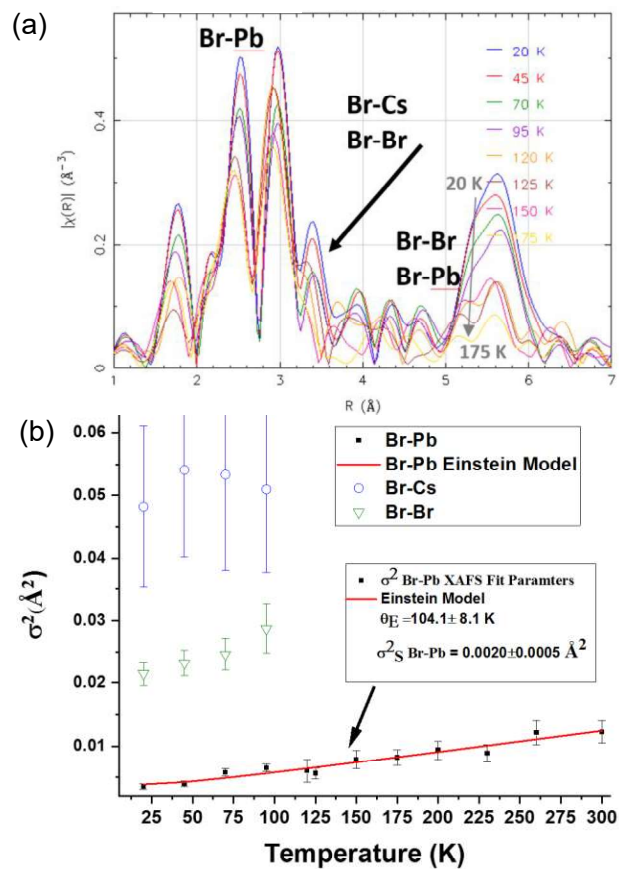

**Fig. S16.** (a) Fourier transform of XAFS data between 20 and 175 K, indicating suppression of high order peaks beyond Pb-Br above ~170 K. Bold labels indicate the peak assignments (e.g., Br-Pb) and light labels indicate the data temperatures values. Three-shell Fits to XAFS data above ~ 95 K are unstable. (b) Extracted XAFS thermal parameters ($\sigma^2$) for the Br-Pb, Br-Cs, and Br-Br atomic pairs for temperatures up to 95 K and for Br-Pb only for higher temperatures. The Br-Pb data was modeled by an Einstein function, yielding a static contribution $\sigma^2_S = 0.0020 \pm 0.0005$ Å$^2$ and Einstein temperature of $\theta_E = 104.1 \pm 8.1$ K.

**(a) 430 K (Im-3 Z=8)**                    **(b) 370 K (P2₁/m Z=8)**

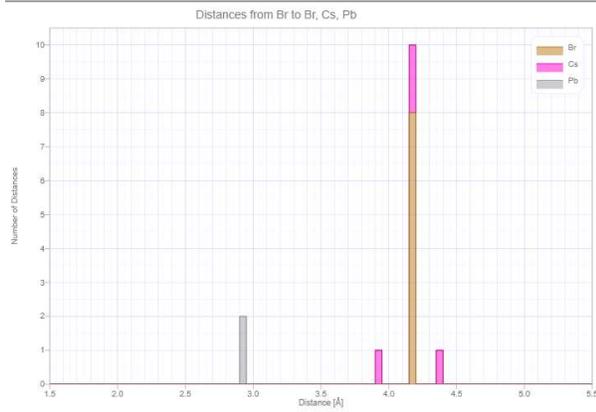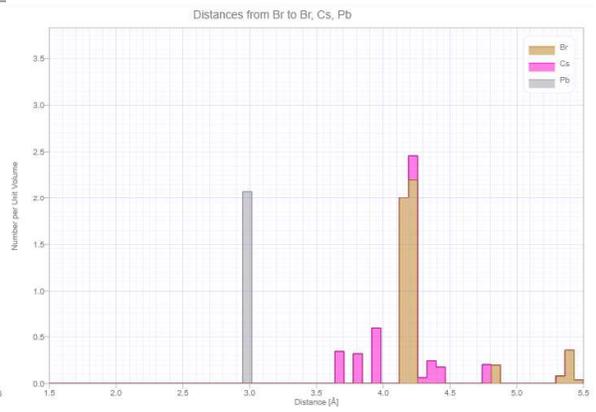

**(c) 230 K (P2₁2₁2₁ Z=4)**                 **(d) 100 K ((P2₁2₁2₁ Z=4)**

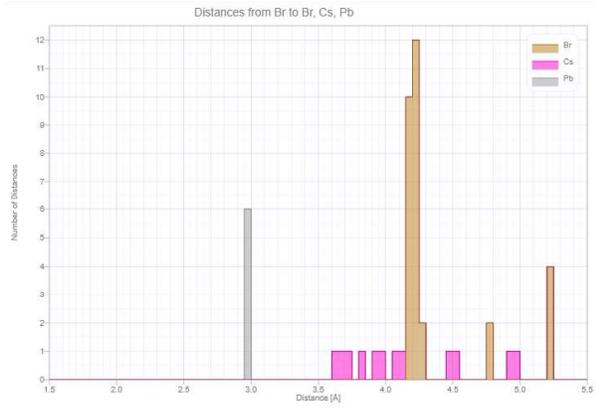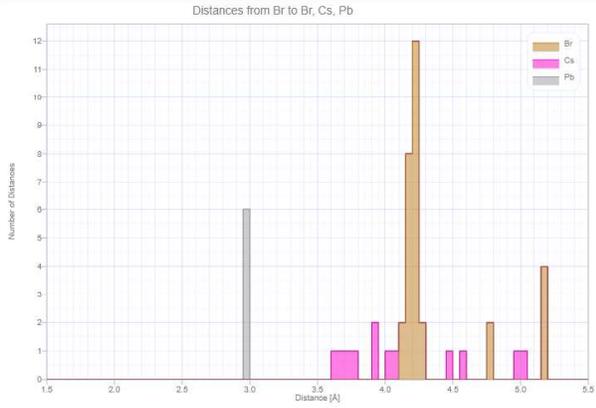

**(e) 230 K (Pnma Z=4)**          **(f) Cubic Structure based on 230 K lattice params. (Pm-3m (Z=1))**

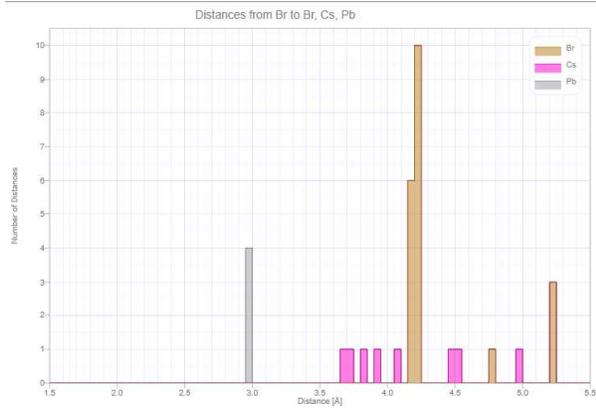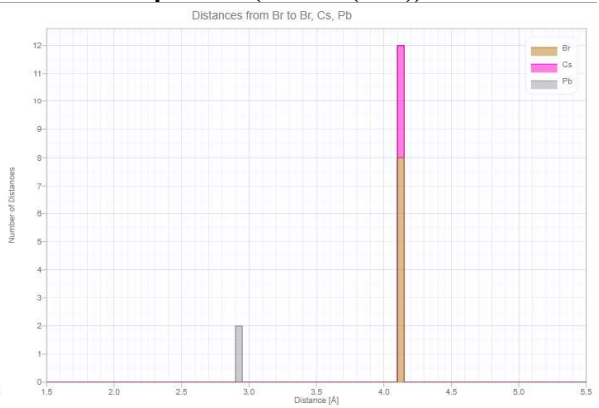

**Fig. S17.** Atomic pair distributions (number of atom-atom distances vs. r) about Br sites derived from single-crystal data for structural solutions at (a) 430 K, (b) 370 K, (c) 230 K, and (d) 230 K. For completeness, the Pnma structure at 230 K is given in (e), and the approximated distribution for a simple cubic cell based on lattice parameters at 230 K is given in (f). Note that in the real sample (not simple cubic structure), the Cs-Br distribution is never a single peak for temperatures up to 450 K, at least. Note also the large spread in the Br-Cs distribution in the $P2_12_12_1$ space group. This spread in positions becomes less broad in the 100 K structure compared to the 230 K structure.

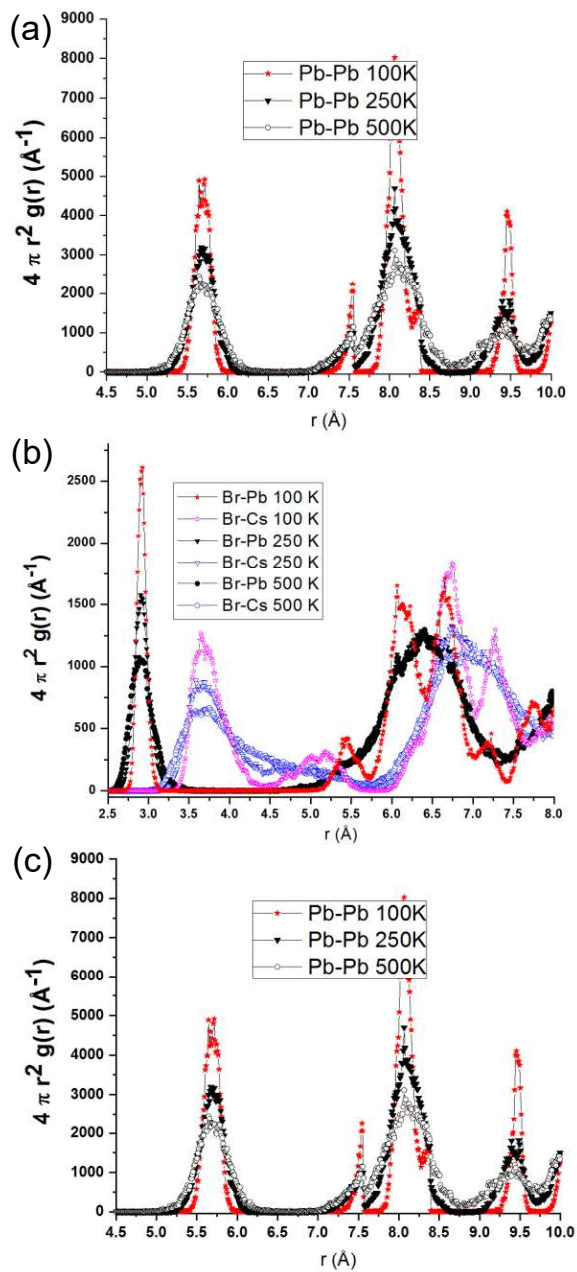

**Fig. S18.** (a) Radial distribution functions for Pb-Pb pairs at 100, 250, and 500 K derived from ab initio molecular dynamics simulations. (b) Corresponding functions for the Br-Pb and Br-Cs pairs. Note the loss of discrete structure and significant broadening occurring on going from 100 to 250 K compared to the smaller changes in going from 250 to 500 K. In panel (c), note that the same abrupt broadening is seen in the Br-Br pair distributions in going from 100 to 250 K. The results are consistent with significant disordering of the Br-Cs and Br-Pb pairs between 100 and 250 K (for increasing temperature).